%% file: main.tex
\documentclass[12pt,a4paper,final]{iopart}

\usepackage{graphicx}
\usepackage{cite}
\usepackage{siunitx}
\usepackage{iopams} 
\usepackage{ulem}

\expandafter\let\csname equation*\endcsname\relax
\expandafter\let\csname endequation*\endcsname\relax
\usepackage{amsmath}
\usepackage{float}
\usepackage{braket}
\include{LCB_macros}

\begin{document}

\title[Network structure of effective models of non-equilibrium quantum transport]{Network structure and dynamics of effective models of non-equilibrium quantum transport}

\author{Abigail N. Poteshman$^{1}$\footnote[2]{Current Address: Committee on Computational and Applied Mathematics, Physical Sciences Division, University of Chicago, Chicago, IL, 60637, USA}, 
Mathieu Ouellet$^{2}$, 
Lee C. Bassett$^2$\footnote[1]{These authors contributed equally.}, 
Danielle S. Bassett$^{1,2,3,4}\dagger$}
\address{$^1$ Department of Physics \& Astronomy, College of Arts \& Sciences, University of Pennsylvania, Philadelphia, PA 19104, USA}
\address{$^2$ Department of Electrical \& Systems Engineering, School of Engineering \& Applied Science, University of Pennsylvania, Philadelphia, PA 19104, USA}
\address{$^3$ Department of Bioengineering, School of Engineering \& Applied Science, University of Pennsylvania, Philadelphia, PA 19104, USA}
\address{$^4$ Santa Fe Institute, Santa Fe, NM 87501, USA}

\ead{\mailto{lbassett@seas.upenn.edu} and \mailto{dsb@seas.upenn.edu}}
\vspace{10pt}
\begin{indented}
\item[]January 2021
\end{indented}

\begin{abstract} 
Across all scales of the physical world, dynamical systems can often be usefully represented as abstract networks that encode the system's units and inter-unit interactions. Understanding how physical rules shape the topological structure of those networks can clarify a system's function and enhance our ability to design, guide, or control its behavior. In the emerging area of quantum network science, a key challenge lies in distinguishing between the topological properties that reflect a system's underlying physics and those that reflect the assumptions of the employed conceptual model. To elucidate and address this challenge, we study networks that represent non-equilibrium quantum-electronic transport through quantum antidot devices --- an example of an open, mesoscopic quantum system. The network representations correspond to two different models of internal antidot states: a single-particle, non-interacting model and an effective model for collective excitations including Coulomb interactions. In these networks, nodes represent accessible energy states and edges represent allowed transitions. We find that both models reflect spin conservation rules in the network topology through bipartiteness and the presence of only even-length cycles. The models diverge, however, in the minimum length of cycle basis elements, in a manner that depends on whether electrons are considered to be distinguishable. Furthermore, the two models reflect spin-conserving relaxation effects differently, as evident in both the degree distribution and the cycle-basis length distribution. Collectively, these observations serve to elucidate the relationship between network structure and physical constraints in quantum-mechanical models. More generally, our approach underscores the utility of network science in understanding the dynamics and control of quantum systems. 
\end{abstract}

%
%
%
%
%

\section{\label{sec:level1:Introduction}Introduction}

The intersection of network science and quantum physics is an emerging area of interdisciplinary research \cite{biamonte2019complex}. Methods from network science have been used to characterize features of quantum networks that are relevant to the design of quantum information and communication systems, such as quantum synchronization \cite{lohe2010quantum}, transport efficiency \cite{mulken2016complex}, and robustness to noise \cite{cabot2018unveiling}. Conversely, quantum effects and dynamics have been applied to complex networks, such as by modeling quantum walks \cite{chakraborty2016spatial} and representing nodes as entangled states \cite{cirac1997quantum}. In both directions of inquiry, this interdisciplinary work has focused on manipulating network structure in order to optimize networks for quantum information processing, storage, and communication technologies. Yet, this focus has necessarily neglected the important space of questions surrounding how network structure emerges naturally and directly from quantum systems themselves. In previous work, we sought to address this gap by considering the structure of mesoscopic quantum networks, and by demonstrating the utility of network characterizations in explaining transport properties \cite{poteshman2019network}. Here we take a complementary approach and ask: What network topology emerges from different physical models of mesoscopic quantum systems? And what do those differences imply about system dynamics and control?

Mesoscopic quantum systems, such as quantum dots and quantum antidots, are of particular interest to those designing quantum information processing devices \cite{loss1998quantum, awschalom2013quantumspintronics,barthelemy2013quantum}.
They are widely tunable and can be efficiently controlled electronically by capacitive coupling to electrostatic gates that can alter their equilibrium charge \cite{kouwenhoven1997introduction,wiel2002electron,hanson2007spins,sim2008electron}. Mesoscopic quantum systems can be probed with transport experiments: electrons tunneling between reservoirs weakly coupled to a mesoscopic system induce transitions between quantum mechanical configurations, whose properties can be deduced from measurements of current and conductance. Features of mesoscopic systems, however, are difficult to characterize and predict since simulating a many-body interacting system is computationally intractable, due to exponential scaling of the system's Hilbert space with particle number \cite{tighineanu2015unraveling, yang2016resonance}. Without true quantum simulators, the best tools available to model mesoscopic systems are numerical, semi-classical models. 

In the recent literature, network science has emerged as a promising tool to offer intuition for the architectures of physical systems that produce mesoscopic dynamics \cite{bagrov2020detecting, valdez2017quantifying, sundar2018complex, zaman2019real}. For example, in a previous study employing a single-particle model of quantum-electronic transport, we demonstrated that statistical characterizations from network science can capture physically-relevant emergent properties of non-equilibrium transport \cite{poteshman2019network}. Yet, the work left unanswered the question of how different physical constraints embedded in various models of quantum phenomena are reflected in the network architecture, thereby informing appropriate control strategies \cite{andrieux2006fluctuation, andrieux2007fluctuation}. That such reflections might exist is intuitively plausible when one considers the nature of the physical models, which can represent quantum states and mechanisms for state excitation quite differently, for example by using different bases or at different levels of approximation \cite{stone1991edge, stone1990schur}. 

\begin{figure*}
\centering
  \includegraphics[width=0.75\textwidth]{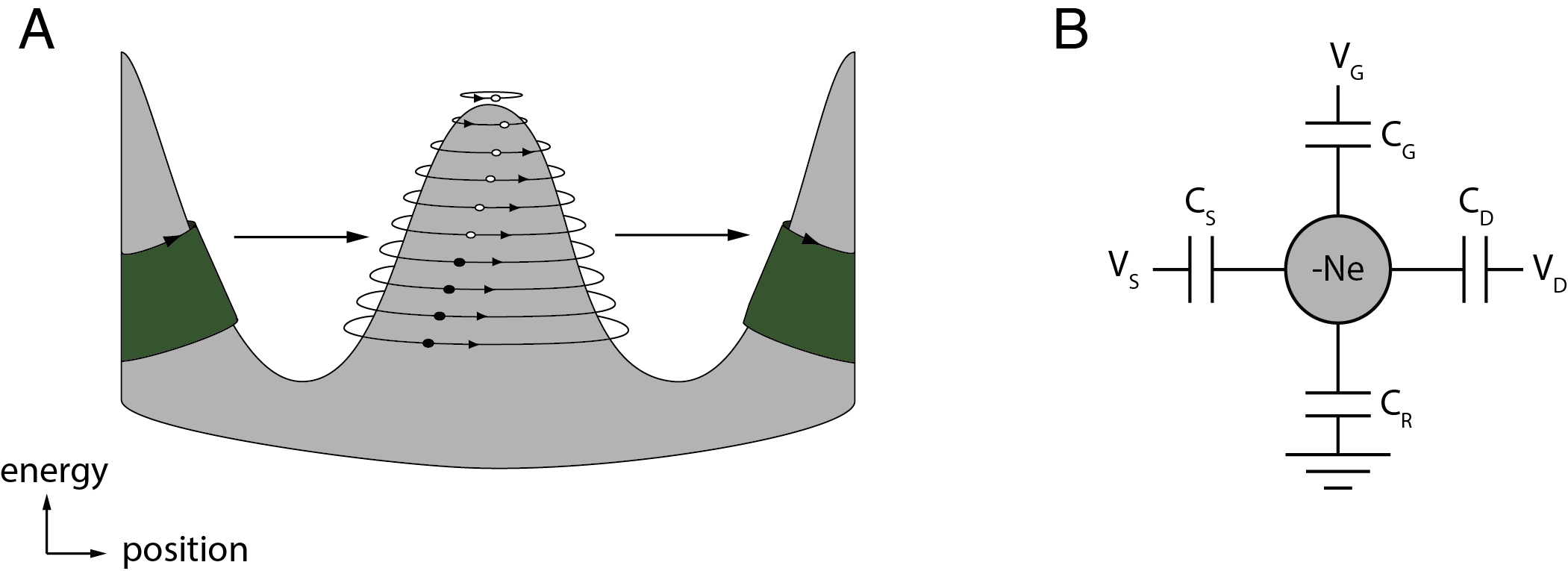}
  \caption{\textbf{An overview of antidot physics.} \textbf{A.} Schematic cross section of an antidot. Tunneling occurs between edge states carrying a sea of electrons (green) and quantized antidot energy states. \textbf{B.} Equivalent capacitor network for the antidot electrostatics. The quantized charge on the antidot is $-Ne$, where $N$ is the number of electrons (relative to a fixed reference configuration) and $e$ is the electron charge. The electrostatic potential of the antidot can be determined by the capacitive couplings ($C_S$, $C_D$, $C_G$) to the source, drain, and gate voltages ($V_S$, $V_D$, $V_G$). Any remaining coupling to other elements of the device is modeled as a capacitive coupling to the ground potential ($C_R$), such that the total capacitance is $C=C_S+C_D+C_G+C_R$. Figure adapted from Ref. \cite{bassett2019probing}.}
  \label{fig:ADdevice}
\end{figure*}

To better understand the interdigitation between physics and topology, we study networks constructed from two models of non-equilibrium transport through a quantum antidot (see Figure \ref{fig:ADdevice}A): a single-particle model and an effective model \cite{bassett2019probing}. Both models produce experimentally accurate time-averaged values of current and conductance from transport experiments, but describe the internal antidot configurations and mechanisms for excitations in different ways \cite{bassett2019probing}. The single-particle model treats quantum states as composed of distinguishable, non-interacting elementary particles, whereas the interacting model describes quantum states in terms of collective quasiparticle excitations of a many-body liquid. That is, the two models are not merely different basis representations of the same physics. We performed a statistical investigation of the models' network topology, paying particular attention to the network's cycle structure and degree distribution, which are high-order and lower-order, respectively, topological characteristics relevant to the propagation of information and control profiles of complex networks \cite{lizier2012information,boccaletti2006complex, campbell2015topological}. In comparing the networks built from these two models, we aim to distinguish the network characteristics that reflect the physically observable phenomena of quantum transport (shared across both models) from those that reflect the underlying physical mechanisms of particular models (differing between models).

\section{\label{sec:level1:Methods}Methods} Here, we extend previous work that built a network model of the energy landscape of non-equilibrium transport through quantum antidots \cite{poteshman2019network}. Expanding upon that study, we now consider an effective model in addition to a single-particle model of antidot states. We also examine the impact of including spin-conserving relaxation effects on network structure. Together, these extensions allow us to ask (and answer) new questions about the relations between physical processes and resultant network topology in mesoscopic quantum systems.

\subsection{\label{sec:level2:ad_transport}Transport regime of interest} 

For an overview of transport through mesoscopic systems, we direct readers to seminal reviews such as Refs. \cite{kouwenhoven1997introduction,wiel2002electron, hanson2007spins, sim2008electron}. Here, we focus on spin-resolved transport through a single quantum antidot at filling factor $\nu_{AD} = 2$ in the integer quantum Hall regime at relatively low magnetic fields ($B < \SI{1}{\tesla}$) (see Figure ~\ref{fig:ADdevice}A). In this regime, both the single-particle and effective models of internal antidot configurations qualitatively describe transport experiments (see Figure \ref{fig:Curr_Cond}A-B and Figure 1A-B in the Supplement) \cite{mace1995general, stone1992edge}. For further details on non-equilibrium transport through quantum antidots in this particular regime, we direct readers to Refs. \cite{poteshman2019network, bassett2019probing}.

To quantitatively characterize transport in an antidot system weakly coupled to extended metallic leads, experiments measure (and models predict) the current, $I$, and the differential conductance, $G =  \frac{dI}{dV}$. Nonzero current indicates the presence of accessible quantum states in the antidot within the energy window defined by the relative electrochemical potentials of the source, $\mu_S$, and drain, $\mu_D$ (see Figure ~\ref{fig:ADdevice}B). Differential conductance reflects changes in the transport configurations, generally due to changes in the alignment of state transitions with $\mu_S$ and $\mu_D$. Differential conductance is typically positive but can become negative is certain configurations. Together, current and conductance are used as both qualitative and quantitative indicators of quantum transport phenomena, and they are the key output of computational models for comparison with experiments \cite{poteshman2019network, bassett2019probing}.

The number of antidot states involved with non-equilbrium quantum transport grows rapidly with the applied bias (see Figure ~\ref{fig:ADdevice}B). The additional states relevant for non-equilibrium transport include excited states that represent different spin and orbital configurations \cite{hanson2007spins}. The ways in which these spin and orbital configurations are connected through tunneling and relaxation events are manifold, leading to a richly structured collective energy landscape. In fact, landscape complexity grows exponentially with particle number; it quickly becomes computationally intractable to calculate transport characteristics analytically. As a result, the best tools available to model non-equilibrium transport through mesoscopic systems are numerical, semi-classical models. Quantum antidot states can be modeled either in terms of the electron state occupation number (see Section \ref{sec:level2:sp_model}), or as edge-waves in the charge distribution (see Section \ref{sec:level2:mf_model}) \cite{stone1992edge}.

\subsection{\label{sec:level2:sp_model} Single-particle model}

Here we provide a brief description of the single-particle model for transport through quantum antidots (see Ref. \cite{poteshman2019network} for further details). The single-particle energies are labeled by orbital ($m=0,1,2,\ldots$) and spin ($\sigma=\pm\frac{1}{2}$) quantum numbers, 
 
\begin{equation}
    \varepsilon_{m\sigma}=m\Delta E_\mathrm{SP}+\sigma E_\mathrm{Z},
\end{equation}

\noindent yielding an energy spectrum of two ladders of equally-spaced energy levels. The spacing between orbital energy levels, $\Delta E_{SP}$, is assumed to be constant, and the separation between energy ladders is the Zeeman energy $E_Z$. Excitations are also governed by these two energy scales, with possible values
  
 \begin{equation}
     E_{ex} = j \Delta E_{SP} \pm q E_Z,
\end{equation} 
 
\noindent where $q = 0$ and $q = 1$ represent spin-conserving or spin-flip transitions, respectively, and $j$ is any integer. Internal quantum states of the antidot are represented as a pair of electronic occupation vectors (\textbf{n\textsubscript{$\uparrow$}}, \textbf{n\textsubscript{$\downarrow$}}), with components $n_{m\sigma} = 0$ or $1$ for each orbital, $m$, and spin, $\sigma$. Since we can track whether electrons occupy specific orbitals, the electrons in the single-particle model are distinguishable. Once the possible electronic occupation vectors are enumerated, a Boolean set of selection rules can be calculated by determining which sets of electronic occupation vectors differ by exactly one electron. That is, if the XOR sum of two sets of electronic occupation vectors is exactly 1, then the transition is allowed. Otherwise, the transition is forbidden.

\subsection{\label{sec:level2:mf_model}Effective model}

We can also consider an effective model of antidot states, in which electrons are indistinguishable and excited states are described as collective excitations of a ``quantum liquid'' around the antidot edge \cite{stone1992edge, macdonald1993quantum}. The effective model is based on the full Hamiltonian for a system of $N$ interacting electrons, within the standard Born-Oppenheimer approximation in which the electronic degrees of freedom are decoupled from those of the lattice \cite{ashcroft1976solid}. The Hamiltonian can be written in the form

\begin{equation}
  \label{equation:Hinteracting}
  \hat{H} = \sum_i^N\HSP_i +
  \frac{e^2}{4\pi\epsilon\epsilon_0}\sum_{i>j}^N\frac{1}{\abs{\mathbf{x}_i-\mathbf{x}_j}},
\end{equation}

\noindent where $\HSP_i$ is the single-particle Hamiltonian acting on the $i^\mathrm{th}$ electron, which is given by

\begin{equation}\label{equation:Hi}
  \hat{h}_i = \frac{1}{2m^\ast}(-i\hbar\mathbf{\nabla}_i+e\mathbf{A})^2
  -e\varphi(\mathbf{x}_i)-g\mu_\mathrm{B}B\hat{s}_{zi}.
\end{equation}

\noindent This general Hamiltonian does not have any known analytic solutions for more than one electron.

Using Hartree-Fock mean-field theory (see Supplement, Section 1.1), we assume that each electron in the multi-electron system is described by its own single-particle wave function (Eq. \ref{equation:Hi}). The multi-electron wave function $\Psi$ can be written as a Slater determinant of orthonomal single-particle spin orbitals, and we can obtain the total energy $E$ for $\Psi$ using the variational principle \cite{bassett2019probing}. In this way, we obtain a 'fermionic' basis of multi-particle states, which characterizes antidot states by occupation numbers of fermion orbitals. Since the fermionic basis states are not general eigenstates of the interacting Hamiltonian (Eq.~\ref{equation:Hinteracting}), we obtained the eigenenergies by diagonalizing the matrix Hamiltonian constructed from the subspace of fermionic basis states with a given $(M,S_z)$, using the rules for addition of angular momentum \cite{stone1992edge}. This process leads to a `bosonic' basis, in which the neutral excitations are described by a spectrum of `edge waves' similar to the one-dimensional Tomonaga-Luttinger liquid model \cite{Luttinger1963,Tomonaga1950}.

The antidot states and transition rules among them are defined as follows. In the effective model the antidot states are given by $\ket{N,S_Z, n_L, n_S}$, where $N$ is the particle number of the state, $S_Z$ is the total spin projection, $n_L \in \mathbb{Z}^+$ is the excitation of the density modes, and $n_S \in \mathbb{Z}^+$ is the excitation of the spin modes. The configuration energy for a state in the interaction model is given by $U(\ket{N, S_Z, n_L, n_S}) = U(\ket{N, S_Z, 0, 0}) + n_L \cdot E_L + n_S \cdot E_S$ where $E_L$ is the energy scale for the density mode excitations and $E_S$ is the energy scale for the spin mode excitations. A transition between two states of the effective model is allowed if $\Delta N = \pm 1$ and  $\Delta n_S \in \{ \frac{1}{2}, -\frac{1}{2} \} - \Delta |S_Z| $. The binary selection rules are weighted by the Clebsch Gordan coefficient for the addition of the spins $S_Z$ and $S_Z \pm \frac{1}{2}$ corresponding to the transition. This model of weighted selections rules qualitatively replicates asymmetries in the conductance map over the voltage space (see Figure \ref{fig:Curr_Cond}B) that are observed in experiments \cite{bassett2019probing}.

\subsection{\label{sec:level2:comp_model}Computational model of transport through a quantum antidot}

The physics of the antidot enters the calculation in the form of a set of quantum states, its associated energy spectrum, and a set of matrix elements for transitions between states. However, the method to construct and solve a rate-equation matrix for the steady-state probabilities of the antidot is agnostic to the physical model used to obtain the quantum states. We used the same master equation approach to obtain steady-state probability occupations for the antidot's configurations as in Ref. \cite{poteshman2019network}.

In Sections ~\ref{sec:level2:sp_model} \& ~\ref{sec:level2:mf_model}, we described two different models of internal antidot configurations, which yield different descriptions of the quantum states and transition rules. In both cases, however, the total particle number, $N$, and the spin projection, $S_z$, are good quantum numbers, and hence the selection rules between quantum states can be written in block-matrix form, e.g.,

\begin{equation}\label{equation:SelRules}
  \begin{pmatrix}
    0 & W^{+\downarrow}_{\Szz-1} & & & \cdots & 0 \\
    W^{-\downarrow}_{\Szz-\frac{3}{2}} & 0 & W^{-\uparrow}_{\Szz-\frac{1}{2}} & & & \vdots \\
     & W^{+\uparrow}_{\Szz-1} & 0 & W^{+\downarrow}_{\Szz}  & & \\
     & & W^{-\downarrow}_{\Szz-\frac{1}{2}} & 0 & W^{-\uparrow}_{\Szz+\frac{1}{2}} & \\
     \vdots & & & W^{+\uparrow}_{\Szz} & 0 & W^{+\downarrow}_{\Szz+1} \\
     0 & \cdots & & & W^{-\downarrow}_{\Szz+\frac{1}{2}} & 0 \\
  \end{pmatrix}.
\end{equation}

\noindent Here, the states are organized as groups that define each block, $\lbrace\ket{N,S_z}\rbrace$, with $S_z$ increasing from top to bottom ($S_{z0}$ is the ground-state spin projection) and $N$ alternating between two adjacent integer values. The sub-matrices $W^{\pm\sigma}_{S_z}$ contain the selection rules for adding ($+$) or removing ($-$) a particle of spin $\sigma$ to a state with initial spin projection $S_z$. 
The specific states included in the model (both the number of blocks and the number of states in each block) are determined through energy and dynamical considerations for a given bias configuration.

%

The transition rates $\gamma_{s' \rightarrow s}$ from antidot configuration $\ket{s'}$ to $\ket{s}$ are calculated according to a combination of antidot selection rules and Fermi's golden rule (see Section 1.2 of the Supplementary Information for a full derivation of the transition rates). The resulting transition rate matrix, $\mathbf{R}$, is defined by $R_{ij} = \gamma_{s_is_j}=\gamma_{s_j \rightarrow s_i}$, where $i$ and $j$ represent different configurations. We seek the steady-state configuration where the total transition rate into and out of each state is equal, and the solution to the master equation yields the steady-state occupation probabilities $\begin{bf}P\end{bf}$ (see Refs. \cite{poteshman2019network,bassett2019probing} for details about the master equation approach). From $\begin{bf}P\end{bf}$, we can compute the current flowing from each spin-polarized reservoir and the spin-resolved conductance \cite{bassett2019probing}.

Using this computational model, we can simulate quantum transport as a function of experimental parameter settings including gate voltages, drain-source bias, magnetic field, and temperature, and we can choose whether to use an energy spectrum based on a single-particle model or an effective model. The settings chosen in this work are motivated by spin-resolved transport experiments in which the underlying physical parameters (\textit{e.g.}, $\Delta E_\mathrm{SP}$, $E_\mathrm{Z}$, and the spin-dependent tunneling rates) have been well characterized \cite{bassett2019probing}. 
Unless indicated otherwise, the temperature is \SI{50}{\milli\kelvin}, $\Delta E_\mathrm{SP}=\SI{30.7}{\micro\electronvolt}$, $E_Z=\SI{45.8}{\micro\electronvolt}$, $E_C=\SI{85}{\micro\electronvolt}$, the effective spin-up tunneling rate is $\SI{500}{\mega\hertz}$, and the effective spin-down tunneling rate is $\SI{50}{\mega\hertz}$. For details on how the experimental parameter settings enter into quantum transport calculations, see Section 1.4 of the Supplementary Materials.

\subsection{\label{sec:level2:relaxation}Spin-conserving relaxation effects}

We can include spin-conserving relaxation effects within each set of antidot configurations with the same number of particles and total spin $\lbrace\ket{N,S_z}\rbrace$ by adding block matrices \textbf{T} describing these processes to the main diagonal of the matrix in eq.~(\ref{equation:SelRules}) \cite{bassett2019probing}. For the single-particle model, we encoded the spin-conserving relaxations as $T_{ij} = 1$ if state $j$ results from moving one of the electrons in state $i$ to the lowest available orbital in state $i$, and $T_{ij} = 0$ otherwise. For the effective model, we set $T_{ij} = 1$ if state $i$ (represented as $\ket{N_i, S_{Zi}, n_{Li}, n_{Si}}$) and state $j$ (represented as $\ket{N_j, S_{Zj}, n_{Lj}, n_{Sj}}$) have the same total spin ($N_i = N_j$) and spin mode excitation ($n_{Si} = n_{Sj}$), and state $i$ is a density mode excitation of state $j$ ($n_{Lj} < n_{Li}$). For both models, the relaxation rate was set at $\SI{500}{\mega\hertz}$ when relxation was included, which is on the same order as the tunneling rates into and out of the antidot.

\begin{figure*}
\centering
  \includegraphics[width=\textwidth]{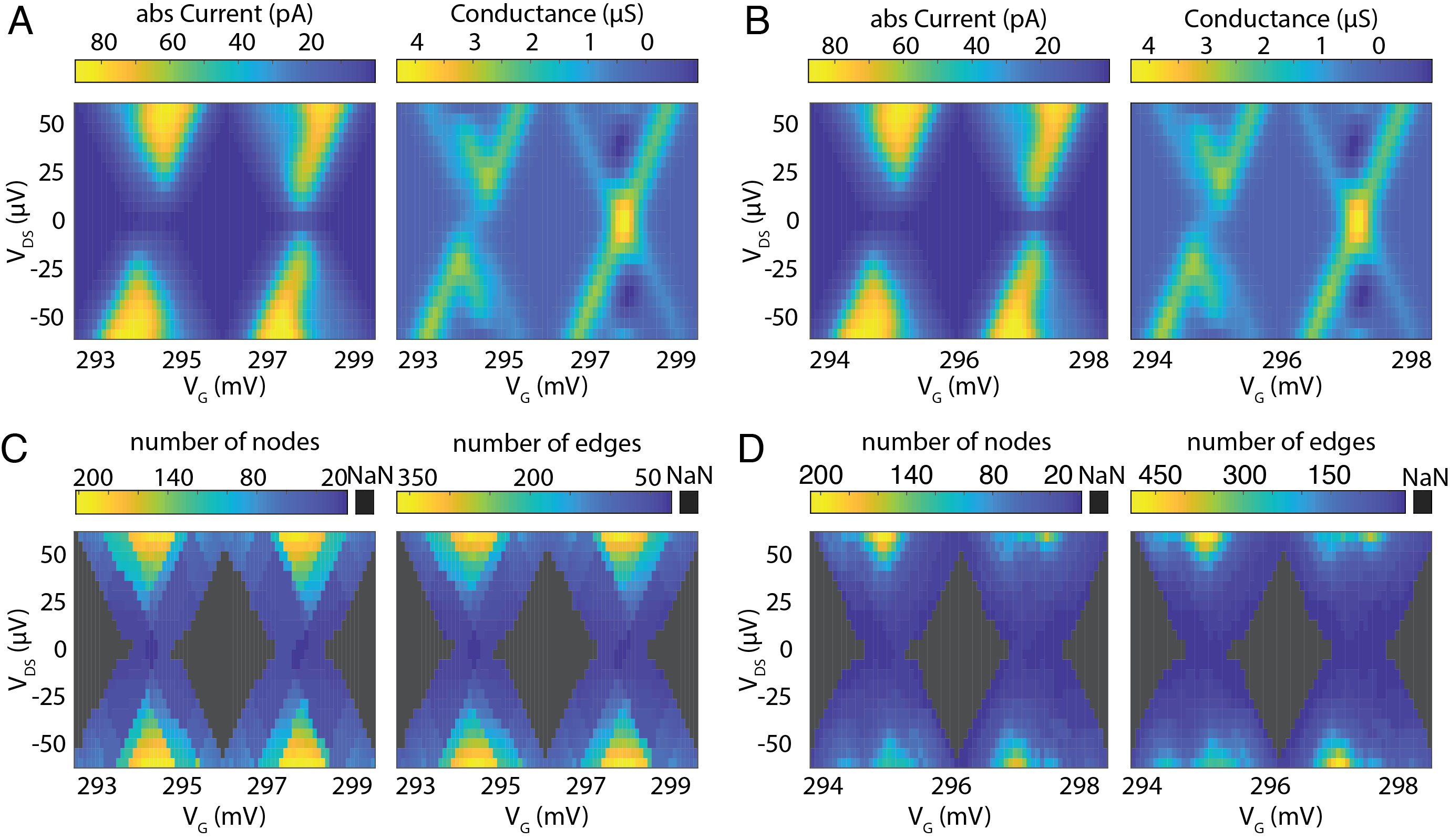}
  \caption{\textbf{Single-particle model versus effective model of non-equilibrium transport through quantum antidots.} Current and conductance calculations based on a single-particle model (\textbf{A}) and an effective model (\textbf{B}) of energy states in quantum antidots. Both models were run with the following parameters: T = $50$ \si{\milli\kelvin}, \textbf{B} = $1.2$ \si{\tesla}, effective spin-up tunneling rate $\Gamma_{\uparrow} = \SI{500}{\mega\hertz}$, and effective spin-down tunneling rate $\Gamma_{\downarrow} = \SI{50}{\mega\hertz}$. All subsequent figures are based on models run with these parameters unless noted otherwise. The number of nodes and the number of edges for the networks corresponding to each set of voltage settings constructed based on a single-particle model (\textbf{C}) and an effective model (\textbf{D}) of energy states. Note that we excluded networks corresponding to voltage settings that result in a current with magnitude less than $1$ \si{\pico\ampere} from our analysis; the values in panels (\textbf{C}) and (\textbf{D}) that are displayed as NaN indicate that networks were excluded. }
  \label{fig:Curr_Cond}
\end{figure*}

\subsection{\label{sec:level2:net_construct}Network construction and statistical characterization}

Similar to constructing a master equation that determines the transition rate matrix and the occupation probabilities, the method to construct networks is agnostic to the underlying physical model used to represent antidot configurations. In the networks representing transport through quantum antidots, the nodes represent antidot configurations and the edges represent possible transitions between configurations after single electron tunneling events and relaxation events. We used the same method to construct networks based on the transition rate matrices \textbf{R} (see Figure ~\ref{fig:AdjMat} A \& B and Figure 2A \& B in the Supplement) and corresponding probability vectors \textbf{P} as reported in Ref. \cite{poteshman2019network}. The thresholding method for the probability vectors is presented in Section 1.5 of the Supplementary Information. 

With these adjacency matrix representations of our network in hand, we can begin to perform rigorous statistical characterizations of network size, density, and topology. The number of nodes $n$ in a network is given by the size of the matrix, and the number of edges in the network is the number of non-zero elements in the matrix (see Figure ~\ref{fig:Curr_Cond}C-D). To evaluate the network's topology, we focus on two statistical measures relevant to a network's capacity to propagate current, signals, nutrients, or other physical, biological, or informational items \cite{lizier2012information,boccaletti2006complex}: the network's degree distribution and cycle structure. The degree of a node $i$ is the sum over $i$ of $A_{ij}$. A narrow degree distribution is indicative of a particularly ordered systems; in a lattice, for example, the degree distribution is a delta function because every node has the same degree, given by the number of its immediately adjacent neighbors \cite{newman2010networks}. A broad degree distribution is indicative of more complexity, where some parts of the system are heavily connected (forming network hubs), and other parts of the system are less connected \cite{albert2002statistical}. In fact, degree heterogeneity is a direct quantification of a network's complexity as formalized in the notion of entropy \cite{lynn2020human}.

The distribution of node degrees is a so-called \emph{lower-order} topological characteristic that considers only the edges in a node's immediate neighborhood: those edges that connect the node directly to other nodes. Ongoing work in the field of network science, however, continues to demonstrate that \emph{higher-order} topological characteristics -- those that characterize the organization of edges which are more than 1 hop away from a node -- have important roles to play in system dynamics and control \cite{lynn2020human, kim2018role, lydon2020hunters, sizemore2018knowledge, sizemore2018cliques}. It therefore seems prudent to consider both lower-order and higher-order topological statistics in our evaluation. In choosing each, we considered the growing body of evidence indicating that degree distribution (lower-order) and cycle structure (higher-order) are two specific network features that consistently shape the dynamics, capacity for information storage, and control profile of a network \cite{albert2002statistical, garcia2012building, garcia2014role, bianconi2003number, lizier2012information}. We therefore complemented the examination of the degree distribution with an examination of the network's cycle basis, which can be algorithmically extracted using the python software package NetworkX \cite{hagberg2008exploring, paton1969algorithm}. 

A cycle is a closed walk that does not retrace any edges immediately after traversing them. Cycles are particularly relevant for understanding transport because a dense cycle structure has been shown to be optimal for transport in the face of spatially and temporally varying loads \cite{katifori2010damage}. A cycle basis is a basis for the vector space of all cycles, defined over $\mathbb{Z}_2$, such that all cycles of a network can be expressed in terms of linear combinations of elements in the cycle basis. Although a network can be decomposed into a cycle basis in many different ways, the length distribution of the cycle basis elements is unique \cite{kavitha2009cycle}. Since extracting the cycle basis of a network can be computed in polynomial time, in contrast with an exhaustive enumeration of all possible cycles which requires exponential time, our analysis of cycle structure is restricted to considering the length of cycle basis elements in this paper \cite{safar2009counting}. Together, the degree distribution and cycle basis allow us to examine the interplay between network topology and mesoscopic physics.

\begin{figure*}
\centering
  \includegraphics[width=\textwidth]{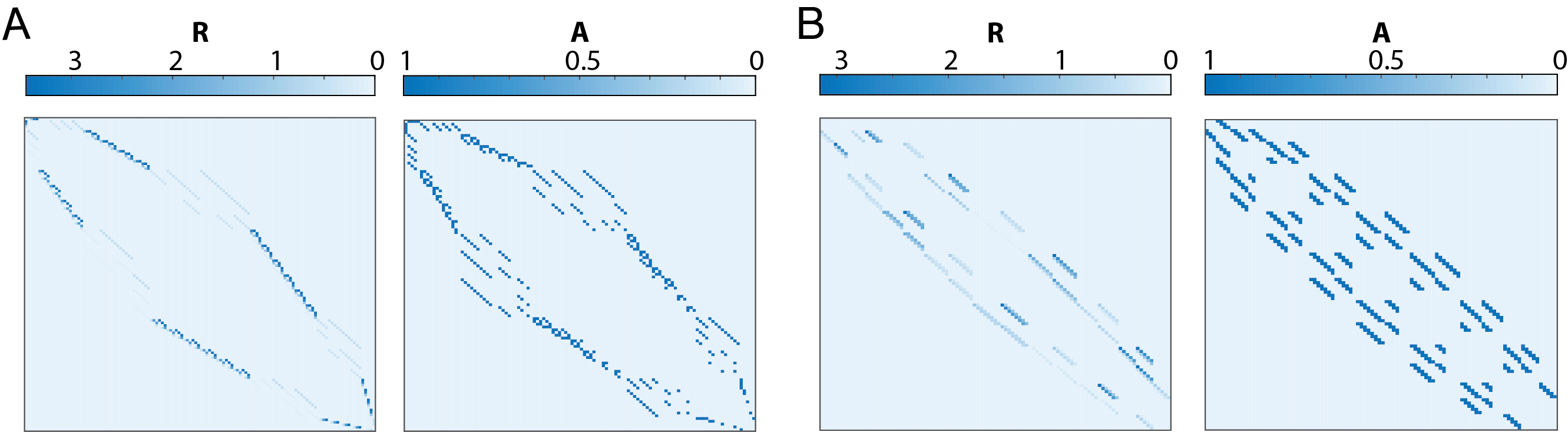}
  \caption{\textbf{Example rate and adjacency matrices.} (\textbf{A}) Example transition rate matrix \textbf{R} (left) and its corresponding adjacency matrix \textbf{A} (right) obtained using a single-particle model of quantum transport. The network these matrices represent corresponds to voltage settings resulting in a current of $\lvert I \rvert$ = $81.0$ \si{\pico\ampere} and 112 states. (\textbf{B}) Example transition rate matrix \textbf{R} (left) and its corresponding adjacency matrix \textbf{A} (right) obtained using an effective model of quantum transport. The network these matrices represent corresponds to voltage settings resulting in a current of $\lvert I \rvert$ = $81.8$ \si{\pico\ampere} and 98 states. }
  \label{fig:AdjMat}
\end{figure*}

\begin{figure*}
\centering
  \includegraphics[width=\textwidth]{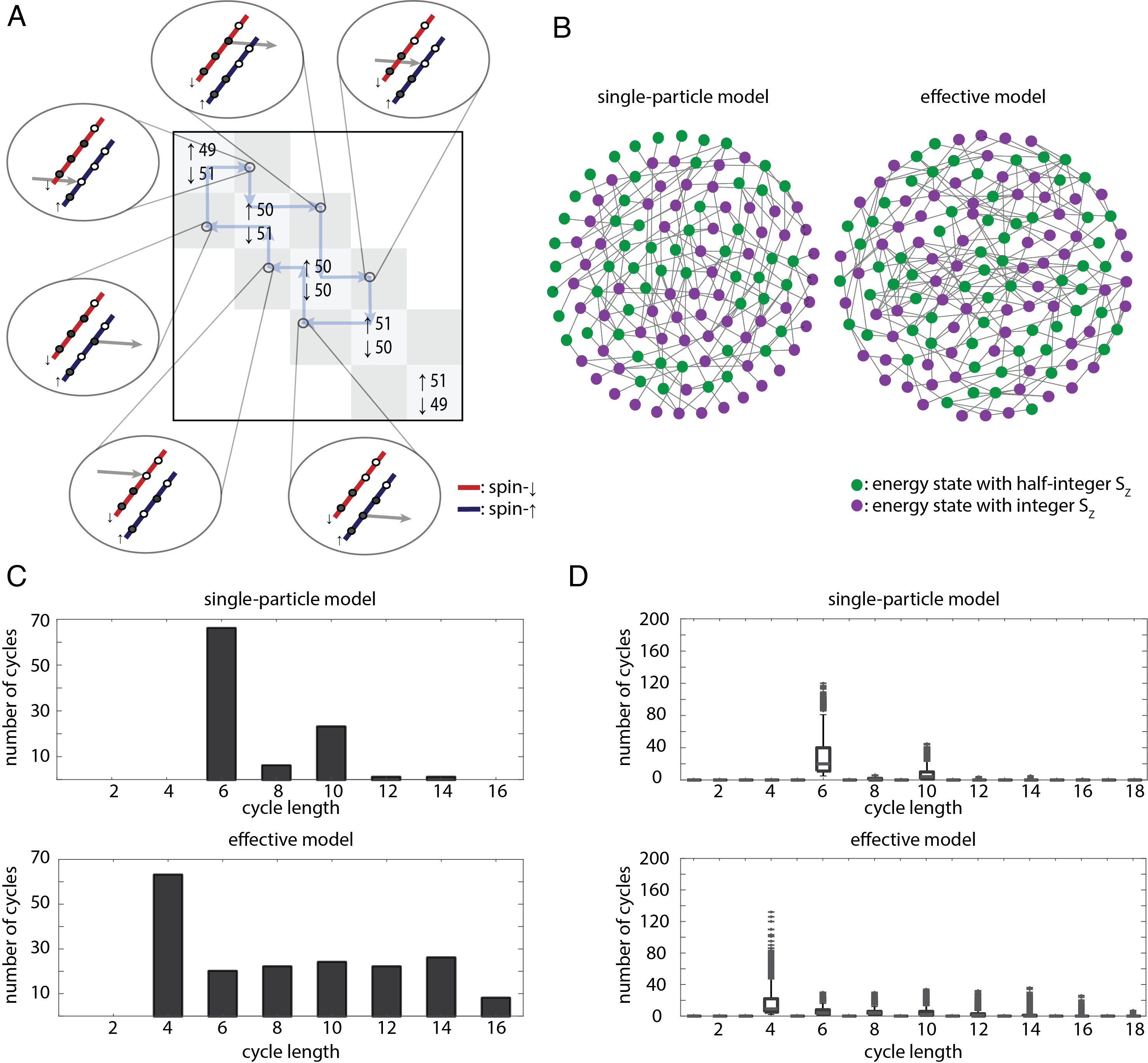}
  \caption{\textbf{Spin constraints result in even-length cycles.} (\textbf{A}) A schematic of cycle trajectory through an adjacency matrix representation of a network. The block diagonal represents antidot energy state configurations that the system may occupy, and the grey off-diagonal blocks store transition rates between antidot states. The cycle shown in the schematic corresponds to a cycle of length 6, where each node in the cycle is a distinct antidot configuration in the diagonal block, and each edge is represented by two blue arrows that pass from one node through a transition rate to a new node. Spin-preserving relaxations occur within a diagonal block. Since the schematic shows an implementation of transport without relaxation, the system must pass through a grey transition state in order to move from one node to another. Since the spin changes by a half-integer amount during each transition, all cycles have an even length. (\textbf{B}) Sample networks with a 2-coloring scheme, where nodes having an integer spin are shown in purple and nodes having a half-integer spin are shown in green for a single particle non-interacting model (left) and for an effective model (right). (\textbf{C}) Distribution of cycle length in the cycle basis space for the single networks shown in panel (\textbf{B}). (\textbf{D}) Distribution of cycle lengths for all networks over the voltage space displayed in Figure \ref{fig:Curr_Cond} corresponding to a current greater than $1$ \si{\pico\ampere} for the networks constructed from the single-particle model (top) and from the effective model (bottom). In each boxplot, the central mark represents the median, the top and bottom edges indicate the third and first quartiles, the whiskers extend to $\pm 2.7 \sigma$, and individual outliers are displayed by '-'.}
  \label{fig:Even_cycles} 
\end{figure*}

\section{\label{sec:level1:Results}Results}

By examining networks constructed from two models of non-equilibrium transport through quantum antidots, we can explore which aspects of energy-state transition networks are common to the physical process of transport versus which aspects reflect particular transport mechanisms. The former will manifest as characteristics common to all transport networks, and the latter will manifest as characteristics that vary across transport models. We begin by examining how spin conservation rules are reflected in the network topology.

\begin{figure*}
\centering
  \includegraphics[width=\textwidth]{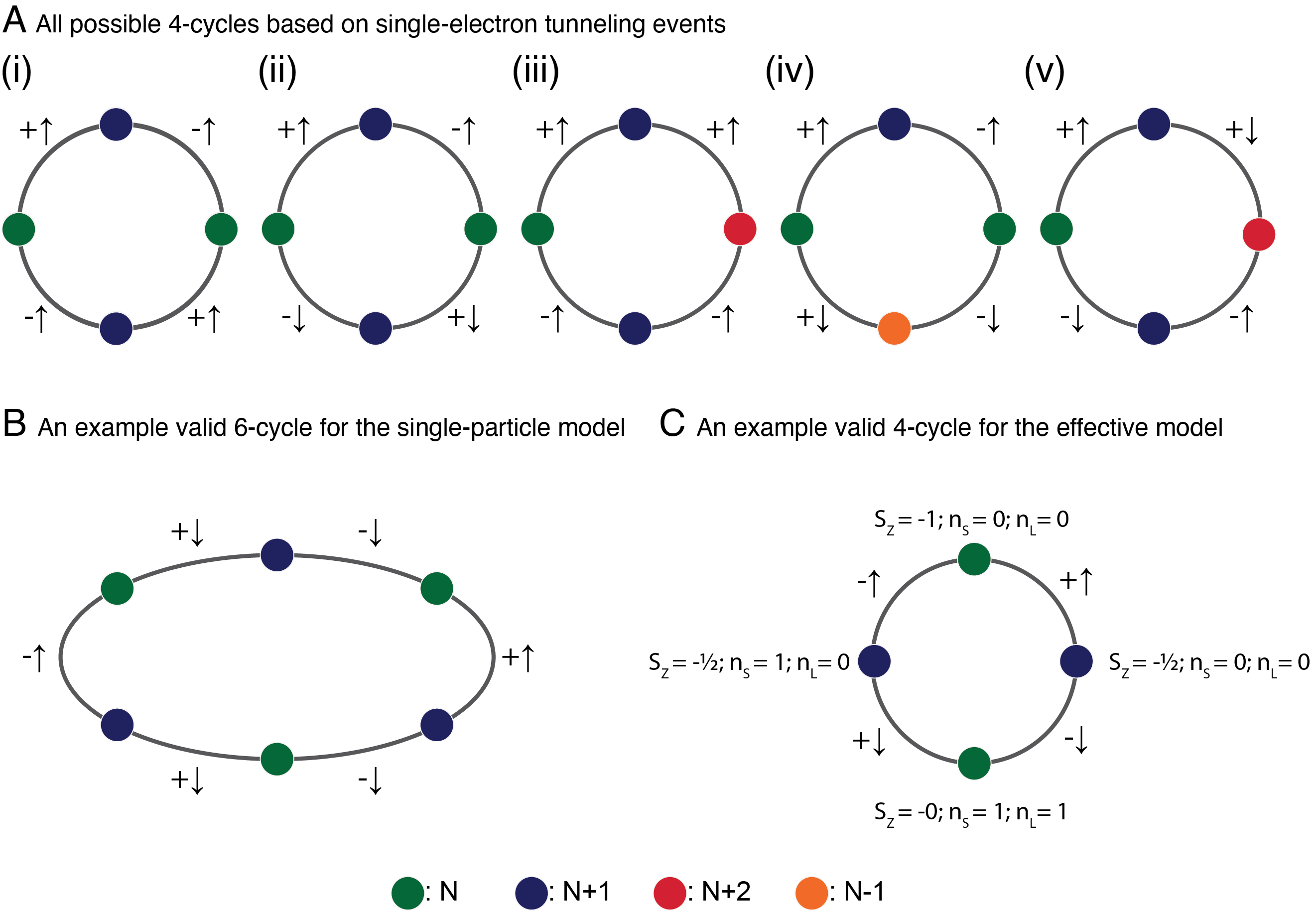}
  \caption{\textbf{Enumeration of cycles.} (\textbf{A}) All possible 4-cycles that involve two electrons tunneling into and two electrons tunneling out of an antidot are shown in schematics \textbf{(i) - (v)}. The color of the nodes indicates the number of electrons in the antidot following the schematics in the clockwise direction; counter-clockwise cycle node labeling is given by changing the sign of the added or subtracted electrons (e.g. $N+1$ in the clockwise labeling becomes $N-1$ in the counter-clockwise labeling). (\textbf{B}) An example of a 6-cycle demonstrating a pattern of electrons tunneling into and out of the antidot that does not invalidate any of the energy or spin conservation rules for the single-particle antidot model. (\textbf{C}) An example of a valid 4-cycle for the effective model with the full quantum numbers labeled for each state. Notice that this cycle is the same as the one shown in subpanel \textbf{A(ii)}.}
  \label{fig:Cycles_4}
\end{figure*}

\subsection{\label{sec:level2:spin_constraints}Spin constraints lead to bipartite networks}

Using two models of non-equilibrium transport through quantum antidots --- a single-particle model and an effective model --- we constructed networks over a range of voltage configurations spanning two Coulomb diamonds (see Figure~\ref{fig:Curr_Cond}A \& B). Transport calculations using both models agree with experimental values of current and conductance (see Figure \ref{fig:Curr_Cond}C \& D), but the models assume quite different transport mechanisms. We first examined networks constructed from both models when excluding relaxation effects. We found that all networks from both models are bipartite. A bipartite network has two classes of nodes and edges that connect only nodes of one class with nodes of the other class \cite{guillaume2004bipartite}. Intuitively then, bipartiteness reflects the shared spin conservation rules that are a common underlying constraint upon both models. Edges represent single tunneling events of electrons into or out of the antidot, so neighboring nodes differ by $\frac{1}{2}$ total spin (see Figure \ref{fig:Even_cycles}A). Understanding edges as single-electron tunneling events leads to a natural 2-color marking scheme of integer versus half-integer total spins (Figure \ref{fig:Even_cycles}B).

As a direct extension of their bipartite nature, we found that all networks contained only even-length cycles in their cycle bases (see Figure \ref{fig:Even_cycles}C \& D). Recall that a cycle is a closed walk that does not retrace any edges immediately after traversing them. To complete a cycle and return to the starting antidot configuration (node), an even number of tunneling events must occur since each tunneling event (edge) changes the total spin of a state (node) by $\frac{1}{2}$. While networks constructed using an effective model have a minimum cycle length of 4 in their cycle bases elements --- an expected minimum length since in the network representations, a 2-cycle is simply an edge --- networks constructed using the single-particle model have a minimum cycle length of 6 (see Figure \ref{fig:Even_cycles}D). 

The difference in minimum cycle lengths stems from a fundamental difference between the single-particle model and the effective model. In the single-particle model, particles are distinguishable, and individual electrons occupy specific excited states. In the effective model, particles are indistinguishable, and excitations of density and spin modes are represented collectively. Enumerating all of the possible 4-cycles, we observe two constraints that prohibit these cycles from occurring for the single-particle model (see Figure \ref{fig:Cycles_4}A), but there exist 6-cycles for the single-particle model that do not violate these constraints (see Figure \ref{fig:Cycles_4}B). First, energy conservation allows the antidot to contain only N and N+1 electrons, eliminating the cycles shown in Figure \ref{fig:Cycles_4}A(iii) - A(v). Note that this constraint also applies to the effective model. Second, an individual electron cannot be added and removed (or removed and added) sequentially, since in the network representation, such a process would constitute two nodes connected by an edge rather than a path through a cycle. This second constraint eliminates the cycles shown in Figure \ref{fig:Cycles_4}A(i) \& \ref{fig:Cycles_4}A(ii) for the single-particle model. This constraint does not apply to the effective model in which electrons are indistinguishable. More precisely, there is no issue in adding or removing the \textit{same} electron since the internal antidot configuration does not track individual electrons, and therefore these 4-cycles are present in networks constructed using the effective model. An example of such a 4-cycle for the effective model is shown in Figure \ref{fig:Cycles_4}C.

\subsection{\label{sec:level2:relaxation_effects}Spin-conserving relaxation effects introduce odd-length cycles}

\begin{figure*}
\centering
  \includegraphics[width=\textwidth]{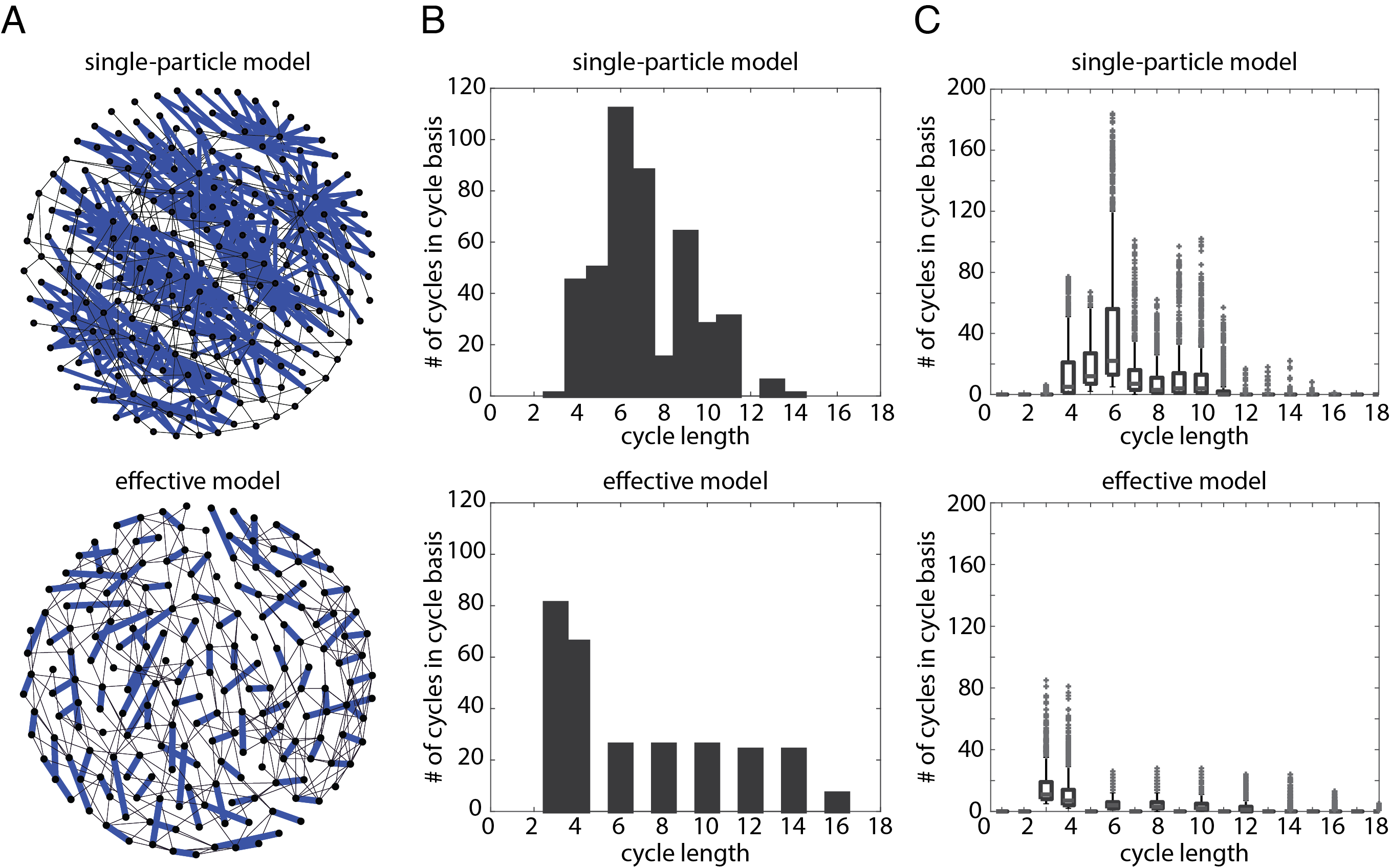}
  \caption{\textbf{Spin-conserving relaxation effects introduce odd-length cycles.} (\textbf{A}) Example networks with edges corresponding to spin-conserving relaxations highlighted in blue for the single-particle model (top) and the effective model (bottom). (\textbf{B}) Cycle length distribution for the single networks shown in panel (\textbf{A}). (\textbf{C}) Distribution of cycle lengths for all networks over the voltage space displayed in Figure \ref{fig:Curr_Cond} corresponding to a current greater than $1$ \si{\pico\ampere} for the networks constructed from the single-particle model (top) and the effective model (bottom). In each boxplot, the central mark represents the median, the top and bottom edges indicate the third and first quartiles, the whiskers extend to $\pm 2.7 \sigma$, and individual outliers are displayed by '-'.}
  \label{fig:Relaxation}
\end{figure*}

As described in Section \ref{sec:level2:relaxation}, both the single-particle model and the effective model can incorporate spin-conserving relaxation effects by adding block matrices to the diagonal of the matrix describing selection rules between eigenstates. For the networks constructed from models that include spin-conserving relaxation effects, an edge represents either a sequential tunneling event or a relaxation event. Including spin-conserving relaxation effects results in a greater number of edges compared with networks excluding relaxation effects corresponding to the same voltage settings (see Supplementary Figure 1), since two nodes may be connected through a spin-conserving relaxation event. Relaxation events do not suppress sequential tunneling events so long as the sequential tunneling rate is on the same order as (or faster than) the relaxation rate; when the relaxation rate is much faster than the tunneling rate, the antidot will effectively remain in its ground state \cite{hanson2007spins}. These relaxation pathways, however, violate the constraints needed to produce a bipartite structure; nodes connected by a relaxation event have the same total spin, so the 2-color marking scheme of integer versus half-integer spins discussed in Section \ref{sec:level2:spin_constraints} no longer holds. 

When we introduce spin-conserving relaxation effects to both models of quantum transport, we observe odd-length cycles; yet, it is important to note that the models' distinct mechanisms of spin-conserving relaxation alter the network structures in different ways (Figure \ref{fig:Relaxation}A). For the single-particle model, a relaxation between antidot configurations $i$ and $j$ is allowed if configuration $j$ results from moving one of the electrons in configuration $i$ to the lowest available orbital in configuration $i$. Since there are multiple possible excited configurations accessible for a given number of electrons, a `ground configuration' node has multiple edges corresponding to relaxations from different possible excited states (Figure \ref{fig:Relaxation}A). As a result, cycles can contain multiple relaxations; cycle basis elements containing an odd number of relaxations result in odd-length cycles in the cycle basis distribution, whereas cycle basis elements containing an even number of relaxations result in even-length cycles (Figure \ref{fig:Relaxation}B-C).  

For the effective model, a relaxation between states $i$ and $j$ is allowed if both configurations have the same number of particles $N$ and the same spin-excitation quantum number $nS$, and if the density spin excitation number of configuration $j$ is less than the density spin excitation number of configuration $i$. Since the only possible density spin excitation numbers are $0$ and $1$, each node has at most one edge corresponding to a spin-conserving relaxation (Figure ~\ref{fig:Relaxation}A). This fact is reflected in the only odd-length cycles in the cycle basis of the effective model as 3-length cycles, since there is at most one relaxation in a cycle basis element (Figure \ref{fig:Relaxation}B-C). In sum, while including spin-conserving relaxation effects results in odd-length elements in the cycle basis for networks constructed using both the single-particle and the effective models, the models' distinct mechanisms to describe excitations yield different cycle structures. 

\begin{figure*}
\centering
  \includegraphics[width=\textwidth]{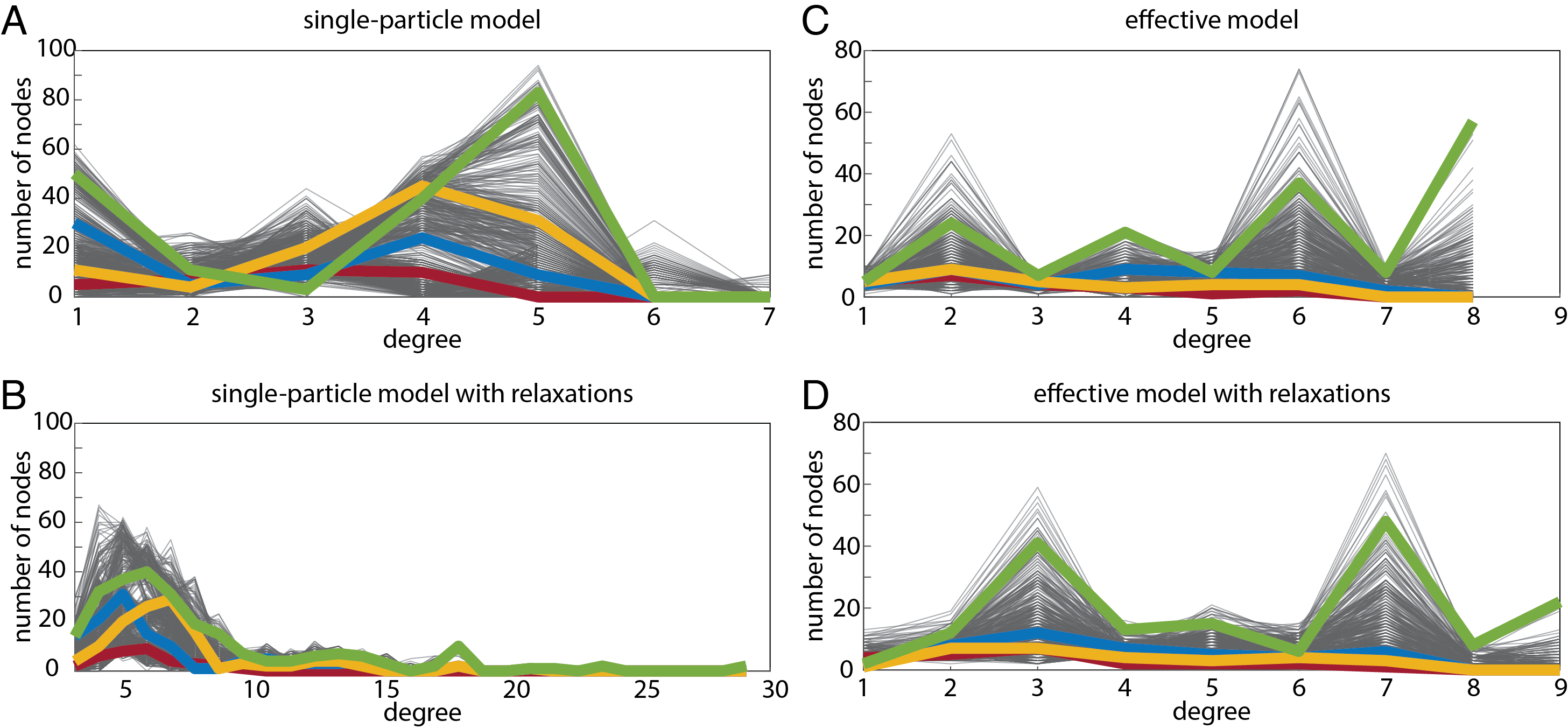}  \caption{\textbf{Physical constraints limit the tail of the degree distribution.} The degree distribution of networks constructed using the (\textbf{A}) single-particle model, (\textbf{B}) single-particle model with spin-conserving relaxation effects, (\textbf{C}) effective model, and (\textbf{D}) effective model with spin-conserving relaxation effects. Each network corresponds to voltage settings resulting in $|I| > \SI{1}{\pico\ampere}$. Networks corresponding to the $25$, $50$, and $75$ percentile values and the maximum value of current over the voltage regime examined are highlighted in red, blue, yellow, and green in panels (\textbf{A}) and (\textbf{C}). The networks corresponding to the same voltage settings in panel (\textbf{B}) as in panel (\textbf{A}) and in panel (\textbf{D}) as in panel (\textbf{C}) are highlighted in the corresponding color.}
  \label{fig:DegDistr}
\end{figure*}

\subsection{\label{sec:level2:deg_distr}Degree distribution}

While cycles are a higher-order topological characteristic that describes the organization of edges which are more than 1 hop away from a given node, degree distribution is a lower-order topological characteristic that concerns the organization of edges directly connected to a given node. While we observed that distinguishability of particles and different mechanisms for relaxation impact the cycle structure of the networks, we seek to understand how the different physical representations of the antidot impact network structure locally. Degree distribution is one of the most fundamental properties of a network used to discern well-known types of networks, such as scale-free networks, Erd\H{o}s-R\'{e}nyi random graphs, and lattices, and it is particularly relevant for questions pertaining to controlling many-body quantum systems with applications in quantum information processing devices, since the distribution of edges in a network governs how information flows locally from one node to another \cite{newman2001random}. More broadly, we are also interested in how the degree distribution of state-transition networks using two different models of internal antidot configurations compares with other naturally-occurring networks. 

Unlike many other social, biological, and physical networks, the quantum transport networks studied here do not exhibit a strongly heavy-tailed degree distribution, nor one following a power law or exponentially truncated power law. This fact is true regardless of which underlying physical model was used to construct the networks and whether spin-conserving relaxation effects were included (see Figure ~\ref{fig:DegDistr}) \cite{barabasi2003scale}. The longest tail observed was for the single-particle model with relaxation. For both models, quantum mechanical selection rules govern how many energy states are accessible from a given energy state, so there is a physical constraint on the maximum number of edges connected to a single node. While the physical constraints on the number of edges limits the length of the distribution's tail, we observe that the degree distributions do not display any scaling behavior (broad scale, scale-free, or single-scale) before the cut-off \cite{amaral2000classes}. Moreover, the distributions are not unimodal, bimodal, or in general smooth. 

Relaxation effects alter edges within a network, so incorporating spin-conserving relaxation effects alters the degree distribution of the network (see Figure ~\ref{fig:DegDistr}). In the effective model, adding spin-conserving relaxation effects results in at most one additional edge for each node; given an antidot configuration with an excited density mode, there is only one possible relaxation to a state with the same total spin and spin excitation but with no density mode. We observe the introduction of spin-conserving relaxations in the effective model as a shift in peaks of the degree distribution between models with and without relaxation (see Figure \ref{fig:DegDistr}C \& D). In the single-particle model, there are multiple possible excited antidot configurations that will relax to the same ground configuration, so including spin-conserving relaxation effects can yield more than one additional edge for a single node. The maximum number of observed edges due to relaxation effects for a single node was 24. We observe the introduction of spin-conserving relaxation effects for the single-particle model through the lengthened tail of the degree distribution (see Figure \ref{fig:DegDistr}A \& B), whereas we observe the introduction of spin-conserving relaxation effects for the effective model as a shift in peaks from even- to odd-degrees in the degree distribution (see Figure \ref{fig:DegDistr}C \& D). Altogether, we observed that spin conservation, particle distinguishability, selection rules, and relaxation mechanisms shape both lower- and higher-order topological characteristics of state-transition networks with implications for the dynamical properties and control profiles of quantum transport networks.

\section{\label{sec:level1:Discussion}Discussion}
\subsection{\label{sec:level2:physics_discussion} Effective models of many-body quantum systems}

Many-body quantum systems are difficult to simulate on classical computing systems, because the number of parameters necessary to represent the system grows exponentially with particle number. As a result, physicists have devised effective models that can be implemented numerically to describe many-body quantum systems. In using effective models, physicists make explicit decisions about how to describe a system as well as which physical effects to include. In the many-body quantum system studied here, chiral edge waves of the quantum antidot in a two-dimensional electron gas system can be viewed either in terms of the electron state occupation numbers (as in the single-particle model) or as edge waves in the charge distribution (as in the effective model) \cite{stone1992edge}. Both models are useful; they yield quantitative values for current and conductance that match experimental data and also recreate many qualitative features of transport experiments. These two models are even connected through the ‘maximum-density-droplet’ state, in which the ground eigenstate of the single-particle model is exactly the ground state solution to the many-particle Hamiltonian, even when accounting for electron-electron interactions \cite{macdonald1993quantum}. 

While the two models studied here both yield values of current and conductance in agreement with time-averaged experimental data, they describe different mechanisms for transport. In the single-particle model, distinguishable electrons tunnel to discrete antidot states labeled by occupation vectors, and individual electrons occupy excited states. In the effective model, indistinguishable electrons tunnel to discrete antidot states labeled by collective spin and density modes. While the difference in how these antidot states are labeled does not impact time-averaged values for current and conductance, our results demonstrate that the choice of underlying physical model impacts the nature of state transitions allowed, which in turn defines the collective energy landscape on which the system exists. These dependencies manifest in two fundamental aspects of the state transition network's structure: the degree distribution and the cycle structure. 

Given the growing interest in network-based approaches to study, model, and control systems governed by quantum physics, our network-based approach to study transport through quantum antidots represents a new paradigm to begin to move toward network-based control strategies \cite{baggio2020data, zanudo2017structure, kim2018role, motter2015networkcontrology}. By finding similarities across networks constructed to represent the same system but using different underlying models, we can better understand which network characteristics reflect physical constraints versus the mechanisms of a specific model. For example, as reported in Section \ref{sec:level2:spin_constraints}, networks representing antidot systems without spin-conserving relaxation effects are bipartite, regardless of which model was used to represent antidot states. This result indicates that control-based strategies designed specifically for bipartite networks may be appropriate to manipulate the antidot to a desired state in regimes in which spin-conserving relaxation effects do not play a role \cite{nacher2013structural}. Altogether, the network-based approach presented here illuminates how different mechanisms to describe transport have different implications for the dynamics and control of many-body quantum systems.

\subsection{\label{sec:level2:cycle_structure_control}Cycle structure and degree distribution in quantum transport networks}

The ability to control many-body quantum systems has become increasingly important in quantum information processing devices, since the ability to identify driver nodes and devise control strategies for quantum systems is crucial in order to store, encode, and process quantum information \cite{cabot2018unveiling, daley2014quantum, ramakrishna1996relation}. Controlling transport is key to many quantum technologies, and both degree distribution and cycle structure impact the control profiles of complex networks \cite{campbell2015topological}. The degree distribution dictates how a signal propagates to neighboring nodes, and cycles give rise to feedback loops. A network is "controllable" if we can identify a set of inputs or driver nodes such that the network can be driven from any initial condition to a target condition within a finite amount of time.


Recent work examining the cycle structure of biological, physical, and engineered systems has shown that cycle structure impacts information propagation, information storage, feedback, and robustness \cite{lind2005cycles, lizier2012information, zhou2018cycle, boccaletti2006complex}. Robustness of a network refers to the ability of a network to continue carrying out a given function when a fraction of its components is damaged. By definition, cycles increase the robustness of transport in a network under an attack that breaks edges, and dense cycle structures have been shown to optimize transport function in response to varying loads and resilience to damage in biological systems such as leaf vasculature and insect wing veins, engineered systems such as city streets, and physical networks such as river deltas \cite{katifori2010damage}. While a tree structure is optimal for transport when there is a spatially and temporally constant load, optimization under varying loads and robustness to an attack dictates the opposite of a tree-like structure \cite{bohn2007structure}. Robustness of transport to an edge attack requires that breaking any one edge does not yield a disconnected network, and while the optimal structure for a network at any single point in time may be tree-like, the overall optimization of spatially and temporally varying transport load requires redundant paths in the form of cycles. Furthermore, cycles play a role in the dynamic stability (the tendency of a network to return to an equilibrium state after a perturbative disturbance) of biological and engineered networks \cite{ma2008ordered}. For example, long cycles are relevant to maintain excitable dynamics in neural networks \cite{garcia2014role}, and short cycles are responsible for maintaining sustained activity \cite{garcia2012building}. Furthermore, dynamic states in networks with longer cycles persist for a longer period of time than those with shorter cycles \cite{woodhouse2016stochastic}. 

Degree distribution impacts the dynamics of how information flows and is distributed throughout a network, as well as controllability and robustness. Dynamically, networks dominated by either even- or odd-degree nodes have different properties: networks dominated by odd-degree nodes tend to noise-induced energy-state changes more than those dominated by even-degree nodes \cite{woodhouse2016stochastic}. Networks with a large fraction of odd-degree nodes have been shown to have more unstable dynamics compared with networks with a higher fraction of even-degree nodes. The reason is intuitive: nodes with an odd degree always have an "unpaired" edge, leaving the network more susceptible to dynamically unstable (chaotic) behavior since "unpaired" edges can cause a network to tend toward getting "stuck" in certain states \cite{woodhouse2016stochastic}. Beyond dynamic stability, the degree distribution of a network has a large role in determining the number of driver nodes in the control profile of a network, where driver nodes tend to avoid high-degree nodes \cite{liu2011controllability}. Degree distribution is also an important topological feature in understanding how robust a network is to both random and targeted attacks, such as an attack in which edges connected to the highest degree nodes are targeted for removal. For example, a bimodal degree distribution optimizes network robustness in the face of both random and targeted attacks \cite{tanizawa2005optimization}.

We found that altering the physical model underlying the description of internal antidot configurations impacts both the degree distribution and cycle structure of network representations. The network structures have different implications for the dynamical properties and control profiles of the antidot system. For example, for the networks constructed using the effective model without spin-conserving relaxation effects, the networks are dominated by even-degree nodes, whereas when spin-conserving relaxation effects are included, the networks are dominated by odd-degree nodes (see Figure \ref{fig:DegDistr}C-D). On a dynamical level, this shift from the relative prevalence of even- versus odd-degree nodes indicates that spin-conserving relaxation effects may serve a destabilizing role in the networks. In terms of network robustness, while the shortest path between two nodes is the most efficient route to communicate information between them, it may not be the optimal route when taking into account traffic, noise, and resistance. In the case of the transport networks, barriers preventing the shortest path between two nodes from being the optimal include passing through an antidot configuration that is energetically costly or improbable to reach \cite{motter2015networkcontrology}. In this case, information may flow through paths that are not topologically the shortest, so short cycles are critical to provide alternative paths and to improve fault tolerance \cite{zhou2018cycle}. We found that in models excluding spin-conserving relaxation effects, the shortest cycle basis elements differ based on whether internal antidot configurations are represented using occupation vectors (as in the single-particle model) or collective spin and density modes (as in the effective model) (see Figure \ref{fig:Even_cycles}). Our results indicate that altering the underlying representation of non-equilibrium transport through quantum antidots has profound implications for understanding the dynamics and devising control strategies for many-body quantum systems.

\subsection{Methodological limitations and future directions}
Several methodological considerations are pertinent to this work. First, we compared the structure of quantum transport networks constructed using two underlying models of internal antidot configurations: a single-particle model and an effective model. The implementation of sequential transport used for both models does not include higher-order co-tunneling processes or spin-flip relaxation effects (due to hyperfine coupling or phonon-mediated spin-orbit interactions \cite{hanson2007spins}), both of which may be present in experiments. Yet, even without including these effects, we find close quantitative agreement between the computational model and experiments \cite{bassett2019probing}. Notably, our approach does allow spin-flip relaxation effects to be incorporated by adding block transition matrices between elements off of the main diagonal of Equation \ref{equation:SelRules}. Second, while the undirected networks allow us to probe the relationship between topology and quantum networks at the most fundamental level, certain physical effects that may impact dynamics are buried. For example, in the undirected networks, both tunneling events and relaxations are represented as undirected edges, when relaxations may be more accurately represented as directed edges. A natural extension of this work would be to explore the network properties of directed networks with edges weighted by occupation probabilities and transitions rates. Third and finally, since enumerating all of the cycles in a network is a computationally-intensive process that typically employs a brute-force depth-first search algorithm \cite{tarjan1972depth}, we explore the cycle basis in order to probe basic features of the cycle structure. While there may be many cycle bases corresponding to a cycle space, the lengths of the cycle basis elements are fixed, so the analysis of length of elements in the cycle bases does not depend on the choice of basis. Future work could further explore the cycle structure of these networks by examining the first Betti number \cite{chung2019exact}, characterizing loop redundancy \cite{sizemore2019importance, blevins2020topology}, and by using other statistical measures to characterize the number of cycles \cite{bianconi2005loops}.

\section{\label{sec:level1:Conclusion}Conclusion}
Using network science to study the energy-state transitions of non-equilibrium transport through a quantum antidot based on two different models of internal antidot states, we demonstrated that structural properties of the network reflect model-specific spin and energy constraints. These constraints result in different minimum-length elements in the cycle bases across models as well as different degree distributions. This understanding of how different physical models of mesoscopic quantum phenomena alters network structure may inform the design and control of quantum devices for quantum simulation, storage, or information processing.

\section{\label{sec:level1:Acknowledgements} Acknowledgements}
ANP acknowledges support from the Benjamin Franklin Scholars Program and University Scholars Program at the University of Pennsylvania. LCB and DSB acknowledge support from the NSF through the University of Pennsylvania Materials Research Science and Engineering Center (MRSEC) DMR-1720530.

\section{References}
\bibliography{ref}
\bibliographystyle{unsrt}

\end{document}


\title[Supplementary Information]{Supplementary Materials for ``Network structure and dynamics paper of effective models of non-equlibrium quantum transport''}

\author{Abigail N. Poteshman$^{1}$\footnote[2]{Current Address: Committee on Computational and Applied Mathematics, Physical Sciences Division, University of Chicago, Chicago, IL, 60637, USA}, 
Mathieu Ouellet$^{2}$, 
Lee C. Bassett$^2$\footnote[1]{These authors contributed equally.}, 
Danielle S. Bassett$^{1,2,3,4}\dagger$}
\address{$^1$ Department of Physics \& Astronomy, College of Arts \& Sciences, University of Pennsylvania, Philadelphia, PA 19104, USA}
\address{$^2$ Department of Electrical \& Systems Engineering, School of Engineering \& Applied Science, University of Pennsylvania, Philadelphia, PA 19104, USA}
\address{$^3$ Department of Bioengineering, School of Engineering \& Applied Science, University of Pennsylvania, Philadelphia, PA 19104, USA}
\address{$^4$ Santa Fe Institute, Santa Fe, NM 87501, USA}

\ead{\mailto{dsb@seas.upenn.edu} and \mailto{lbassett@seas.upenn.edu}}

\newpage



\subsection*{Summary of the Contents of this Supplement}
In this supplementary materials document, we provide additional methodological details and supplementary results that support the findings reported in the main text. We divide the document into two main sections: Supplementary Methods and Supplementary Results. In the Supplementary Methods section, we begin by describing the calculations that we performed to obtain a fermionic basis for the effective model. Next we describe calculations to estimate transition rates for the master equation. Then we describe the calculations that we performed to estimate current and conductance based on the occupation probabilities obtained from the master equation. Finally, we describe the rational for our choice of threshold value for the transition probability matrices. Next we turn to a Supplementary Results section where we report the results for the current, conductance, and size of networks for models including spin-conserving relaxation effects. Then we give sample adjacency and rate matrices for networks including spin-conserving relaxation effects. Finally, we report results from degree-distribution preserving random matrices that demonstrate that the cycle basis is independent from the degree distribution in these quantum transport networks.

\section{Supplementary Methods}
\label{sec:suppl_methods}

Sections \ref{subsection:interacting_model} - \ref{sec:CalcIG} were adapted from Ref. \cite{bassett2019probing}.

\subsection{Obtaining a fermionic basis for the effective model}
\label{subsection:interacting_model}

The Hartree-Fock (HF) method is one of several `mean field'
approaches to this problem, in which each electron in a system is
influenced by an effective potential due to the charge density of
all the other electrons. In particular, we assume that each electron
in the system is described by its own single-particle wave function, such that
the multielectron wave function may be written as a Slater
determinant of orthonormal SP spin orbitals $\psi_i$:
\begin{equation}
  \label{eq:SlaterDet}
  \Psi =\frac{1}{\sqrt{N!}}
      \begin{vmatrix}
            \psi_1(\xi_1) & \psi_1(\xi_2) & \cdots & \psi_1(\xi_N) \\
            \psi_2(\xi_1) & \psi_2(\xi_2) & \cdots & \psi_2(\xi_N) \\
            \vdots & \vdots & \ddots & \vdots \\
            \psi_N(\xi_1) & \psi_N(\xi_2) & \cdots & \psi_N(\xi_N)
      \end{vmatrix},
\end{equation}
where $\xi_i$ represents both the position
and spin projection of the $i^\mathrm{th}$ particle.\footnote{%
It is assumed that the orbital and spin parts of the wave function
are separable, i.e.\ that $\psi_i(\xi) =
\psi_{n_i,m_i}(\mathbf{x})\chi_i(s)$, where $\chi=\bigl(
\begin{smallmatrix}
  1 \\ 0
\end{smallmatrix}\bigr)$
or $\bigl(
\begin{smallmatrix}
  0 \\ 1
\end{smallmatrix}\bigr)$ in terms of the argument $s=1,2$ in spin
space. Matrix elements between spin orbitals therefore imply
integration of spatial coordinates and summation
over spin projections.} %
By construction, this wave function satisfies the antisymmetry
requirement for fermions, that is
\begin{equation}\label{eq:Pauli}
  \Psi(\xi_1,\xi_2,\ldots,\xi_i,\ldots,\xi_j,\ldots,\xi_N) =
      -\Psi(\xi_1,\xi_2,\ldots,\xi_j,\ldots,\xi_i,\ldots,\xi_N),
\end{equation}
since the interchange of particles $i$ and $j$ corresponds to the
interchange of two columns of the determinant, and hence a change of sign. Thus the Pauli exclusion principle is satisfied: the wave
function $\Psi$ vanishes when $\xi_i=\xi_j$ for any $i\neq
j$.\footnote{%
For similar reasons, clearly $\Psi=0$ identically if any
$\psi_i=\psi_j$ for $i\neq j$, enforcing the condition that the SP
spin orbitals chosen must be distinct.}

General expressions for matrix elements between determinantal wave
functions like \eqnref{eq:SlaterDet} are well known, given for
example in Ref. \cite{bethe1986intermediate}. In the
expectation value for the energy of $\Psi$, many of the terms vanish due to the orthogonality of the $\psi_i$, leaving
\begin{equation}
  \langle \Psi \lvert \hat{H}\rvert\Psi\rangle =
      \sum_i\langle i\lvert\hat{h}\rvert i\rangle +
      \sum_{i<j}\Bigl(\langle ij\lvert V_\mathrm{C}\rvert ij\rangle -
          \langle ij\lvert V_\mathrm{C}\rvert ji\rangle\Bigr),
\end{equation}
in terms of the SP energies
\begin{equation}
  \langle i\rvert\hat{h}\lvert i\rangle = \int d\bx\,
      \psi_i(\bx)\hat{h}_i\psi_i(\bx) = \esp_i,
\end{equation}
and two-particle matrix elements of the Coulomb operator
\begin{equation}
  \langle ij\lvert V_\mathrm{C}\rvert kl\rangle  =
      \frac{e^2}{4\pi\epsilon\epsilon_0}
      \delta_{\sigma_i\sigma_k}\delta_{\sigma_j\sigma_l}
      \iint d\bx d\bx^\prime\,
      \psi_i^\ast(\bx)\psi_j^\ast(\bx^\prime)
      \frac{1}{\abs{\bx-\bx^\prime}}
      \psi_k(\bx)\psi_l(\bx^\prime),
\end{equation}
where we have completed the spin summations using the identity
\begin{equation}
  \sum_s\chi_i^\dagger(s)\chi_j(s) = \delta_{\sigma_i\sigma_j}.
\end{equation}
The total energy of the state $\Psi$ may therefore be written in the form\footnote{%
We have used the facts that $J$ and $K$ are symmetric in $i,j$ and
that $J_{ii}=K_{ii}$ in rewriting the sum in \eqnref{eq:HFEtot}.} %
\begin{equation}\label{eq:HFEtot}
  E = \sum_i\esp_i +
  \frac{1}{2}\sum_{ij}(J_{ij}-\delta_{\sigma_i\sigma_j}K_{ij}),
\end{equation}
where
\begin{eqnarray}
  J_{ij} & = & \frac{e^2}{4\pi\epsilon\epsilon_0}
      \iint d\bx d\bx^\prime\,
      \abs{\psi_i(\bx)}^2
      \frac{1}{\abs{\bx-\bx^\prime}}
      \abs{\psi_j(\bx^\prime)}^2, \label{eq:Jij}\\
  K_{ij} & = & \frac{e^2}{4\pi\epsilon\epsilon_0}
      \iint d\bx d\bx^\prime\,
      \psi_i^\ast(\bx)\psi_j^\ast(\bx^\prime)
      \frac{1}{\abs{\bx-\bx^\prime}}
      \psi_j(\bx)\psi_i(\bx^\prime) \label{eq:Kij}
\end{eqnarray}
are the so-called `direct' and `exchange' Coulomb matrix elements,
respectively.

By the variational principle, the energy $E$ thus obtained for our
choice of wave function $\Psi$ represents an upper bound on the
ground-state energy of the system. The best approximation of a
single Slater determinant like \eqnref{eq:SlaterDet} to the true
ground state of the system can therefore be found from the
variational condition $\delta\langle \hat{H}\rangle=0$ under
arbitrary variations $\delta\psi_i$.  Using the method of Lagrange
multipliers to enforce the condition that the $\psi_i$ are
normalised,\footnote{%
It can be shown that the $\psi_i$ remain orthogonal as a result of
this calculation \cite{jackiw2018intermediate}.} %
this may be written
\begin{equation}
  \frac{\delta}{\delta\psi_i}\left[
      \langle\Psi\rvert\hat{H}\lvert\Psi\rangle
      + \sum_i\varepsilon_i\left(\int\abs{\psi_i(\bx)}^2d\bx - 1\right)
  \right] = 0.
\end{equation}
This procedure leads to the Hartree-Fock equations
\begin{equation}
  \label{eq:HFeqn}
  \begin{split}
  \varepsilon_i\psi_i(\bx)  = \hat{h}\psi_i(\bx)
      & +\sum_j\int d\bx^\prime
          \frac{\abs{\psi_j(\bx^\prime)}^2}{\abs{\bx-\bx^\prime}}\psi_i(\bx)\\
      & \qquad
      -\sum_i\delta_{\sigma_i\sigma_j}\int d\bx^\prime
          \frac{\psi_j^\ast(\bx^\prime)\psi_i(\bx^\prime)}{\abs{\bx-\bx^\prime}}
          \psi_j(\bx),
  \end{split}
\end{equation}
which resemble a set of SP Schr\"{o}dinger equations. Indeed, by
taking the inner product of \eqnref{eq:HFeqn} with
$\psi_i^\ast(\bx)$, we see that the eigenvalue $\varepsilon_i$
represents the part of the total energy $E$ involving the
$i^\mathrm{th}$ electron:
\begin{equation}\label{eq:HFei}
  \varepsilon_i =
  \esp_i+\sum_j(J_{ij}-\delta_{\sigma_i\sigma_j}K_{ij}).
\end{equation}

The fermionic basis states discussed thus far are in general not eigenstates of the interacting Hamiltonian (Equation 3 in the main text), because the SP angular momentum operators $\hat{L}_{zi}$, acting on the $i^\mathrm{th}$ electron only, do not commute with the electron-electron interaction term.
The total system is still rotationally symmetric, however, and so the total angular momentum projection $M = \sum_{m\sigma} mn^\mathrm{h}_{m\sigma}$ is a good quantum number of the multiparticle state. Similarly, although the individual spin operators $\hat{s}_{zi}$ do commute with the Hamiltonian, our choice of a Slater-determinant wave function couples the individual spins to the spatial symmetry (and hence the energy) of the state.  Hence
we must consider instead the total spin projection $S_z = \frac{1}{2}\sum_m
(n^\mathrm{e}_{m\uparrow}-n^e_{m\downarrow})$.

\subsection{Calculating transition rates for the master equation system\label{sec:MasterEqn}}

We describe the quantum mechanics of an antidot using a set of `fermionic' states $\lvert s\rangle =
\lvert \lbrace n_{\ell,\sigma}\rbrace\rangle$, labeled by occupation numbers $n_{\ell,\sigma}=0,1$ for the state with orbital and spin quantum numbers $\ell$ and $\sigma$. We split the system into three parts such that the total Hamiltonian $H = H_\mathrm{antidot} + H_\mathrm{res} + H_\mathrm{tun}$ represents the physics of the antidot, the reservoirs, and the tunneling between them, respectively. The Hamiltonian for the antidot is
\begin{equation}
  H_\mathrm{antidot} = \sum_s E_s \lvert s\rangle\langle s\rvert,
\end{equation}
where, within the constant interaction model,
\begin{equation}
  E_s = \sum_{\ell\sigma}\varepsilon_{\ell\sigma}n_{\ell\sigma} +
  \frac{1}{2}\Ec(N-\nG)^2.
\end{equation}
Here, $\Ec = e^2/C$ is the charging energy determined using the equivalent capacitor network (see Fig.~1C in the main text), $N$ is the number of electrons in the antidot, and $\varepsilon_{\ell\sigma}$ represents the single-particle eigenenergies of the antidot.

Similarly, the Hamiltonians describing the reservoirs and tunneling to and from the antidot are given in second-quantized form by
\begin{subequations}
\begin{align}
  H_\mathrm{res} & = \sum_{r = \mathrm{S,D}}\left[
      \sum_{k\sigma}\varepsilon_\ksr a_\ksr^\dag a_\ksr + \mu_r
      \hat{n}_r \right], \\
  H_\mathrm{tun} & = \sum_{r=\mathrm{S,D}}\left[\sum_\kls
      T^r_\kls a^\dag_\ksr a_\ls + \text{h.c.}\right],
\end{align}
\end{subequations}
where the leads (assumed to be non-interacting) are labeled by reservoir $r$, wave vector $k$, and spin $\sigma$. The operators $a_\ksr$ and $a_\ls$ annihilate particles in the lead states $\lvert k\sigma\rangle$ of reservoir $r$ and antidot states $\lvert\ls\rangle$, respectively, and $\hat{n}_r$ is the particle-number operator for lead $r$, with chemical potential $\mu_r= -eV_r$.

Assuming that the couplings to the leads $T^r_\kls$ are small relative to the thermal energy, $k_\mathrm{B}T$, such that thermal fluctuations dominate over quantum-mechanical fluctuations, we can use Fermi's golden rule to write the tunneling rates for the transition between antidot states $\sptos$ and reservoir states $\kptok$ to first order as
\begin{equation}\label{eq:FGRrate}
  W^p_{s^\prime\chi^\prime\rightarrow s\chi}\simeq
      \frac{2\pi}{\hbar}\Bigl\lvert\langle \chi s\rvert H_\mathrm{tun}
          \lvert \chi^\prime s^\prime\rangle\Bigr\rvert^2
      \delta(E_s - E_{s^\prime} + E_\chi - E_{\chi^\prime}+p\mu_r),
\end{equation}
where $p = \pm1$ denotes the change of electron number on the dot, and $E_\chi$ is the energy of the reservoir state $\chi$. We are interested in the rates between individual antidot states, which are obtained by summing out the contributions from all lead states,
\begin{equation}\label{eq:sumoutleads}
  \gamma^p_\sptos = \mspace{-18mu} \sum_{\substack{\chi\chi^\prime \\ N(\chi^\prime) =
  N(\chi)+p}}
      \mspace{-18mu} W^p_{s^\prime\chi^\prime\rightarrow s\chi}
      \rho^\mathrm{eq}_\mathrm{res}(\chi^\prime),
\end{equation}
where $\rho^\mathrm{eq}_\mathrm{res}$ is the equilibrium density of states in the reservoirs. We obtain the result
\begin{subequations}\label{eq:gammapm}
  \begin{align}
    \gamma^+_{r,\sptos} & = \sum_\llps \Gamma^r_\llps(E_s-E_\sp)
        \bra{s}a^\dag_\ls\ket{\sp} \bra{\sp} a_\lps \ket{s}
        f_r(E_s - E_\sp), \label{eq:gammap}\\
    \gamma^-_{r,\sptos} & = \sum_\llps \Gamma^r_\llps(E_\sp - E_s)
        \bra{s}a_\ls\ket{\sp} \bra{\sp}a^\dag_\lps\ket{s}
        \Bigl[1-f_r(E_\sp - E_s)\Bigr],\label{eq:gammam}
  \end{align}
\end{subequations}
where the spectral function is defined as
\begin{equation}\label{eq:spectralfn}
  \Gamma^r_\llps(E) = \frac{2\pi}{\hbar}\sum_k T^r_\kls
  T^{r\ast}_{k\ell^\prime\sigma} \delta(E-\varepsilon_\ksr),
\end{equation}
and
\begin{equation}
  f_r(E) = \frac{1}{1+e^{(E-\mu_r)/k_\mathrm{B}T}}
\end{equation}
are the Fermi functions that describe the occupation of states in the reservoirs.

\subsection{Model of spin-dependent nonlinear transport \label{sec:TransportModel}}

For a given experimental configuration\,---\,consisting of a set of tunnel barriers (possibly spin- and/or energy-dependent) and bias (possibly mode-dependent through the non-equilibrium population of edge modes)\,---\,we have a `shell' within which we can explore different physical models for the antidot. Given a set of values for the external fields (gate voltages, magnetic field, and drain-source bias), we can determine the ground-state configuration, but if the energy spacing between states is small, or if $\Vds$ is large, the steady-state solution will contain significant populations of many excited states as well. To determine which subset of excited states participates in transport, we start with a relatively small subset of states, chosen to be the ground-state configuration plus all of the excited states that are accessible through a single tunneling event, i.e., the states with energy $\varepsilon_i$ such that the chemical potential
\begin{equation}
  \mu_i = \varepsilon_i - \varepsilon_\mathrm{GS}
\end{equation}
is within the energy window defined by the chemical potentials of the leads:
\begin{equation}
\bigl[\min(\muS,\muD)-\Etherm,\,\max(\muS,\muD)+\Etherm\bigr],
\end{equation}
where $\Etherm\approx 4kT$. If we find that many of these excited states have significant occupation probabilities, we can add to this set all of the states that are connected to the significant excited states through the same rule, in terms of the chemical potentials for transitions from each excited state. By continuing to expand the set of states in this way, we will eventually reach a situation in which all of the newly-added states have sufficiently low occupation probability for convergence of the transport current to a desired tolerance. In the results reported here and in the main text, the occupation probability threshold was $1 \times 10^{-8}$.\\

\indent To calculate the spin-dependent current, we need a method of organizing the antidot configurations that allows us to keep track of the \emph{spin} of each electron which tunnels into or out of the antidot. Here we outline the procedure that we use to accomplish this, using the fermionic configurations defined by occupation vectors $(\nupvec,\ndnvec)$ as an example. To begin, we consider only transitions between ground-state configurations with different occupation numbers $N$ at zero bias, with chemical potentials $\mu_0(N)$. Given a set of capacitances as described in Sec. IIA in the main text, the condition $\mu_0(N)=0$ defines the value of the gate voltage $\Vg$ at which charge degeneracy occurs for the $N\leftrightarrow\Npone$ transition at zero bias. In between these resonance positions, the condition $\mu(\Nmone) = -\mu(N)$ defines an inflection point within each Coulomb blockade region. On one side of the inflection point we need only consider configurations with occupation numbers $(\Nmone,N)$, while on the other we consider only $(N,\Npone)$ states. In the plane of $(\Vg,\Vds)$, these become inflection lines that pass vertically through the center of each Coulomb diamond, and divide the calculation region by the occupation numbers involved.

Next, we divide the configurations within each region (defined by occupation numbers $N$, $\Npone$) by their total spin projection $S_z$. Suppose the ground-state spin for the $N$-particle state is $\Szz$ and for the $\Npone$ particle state is $\Szz-\frac{1}{2}$. Given these values, we begin by constructing the vector of configurations:
\begin{equation}
    \lbrace{\ket{\Psi_\mathrm{AD}}\rbrace} =
  \begin{pmatrix}
    \lbrace\ket{\Npone,\Szz\!-\!\frac{3}{2}}\rbrace \\
    \lbrace\ket{N,\Szz\!-\!1}\rbrace \\
    \lbrace\ket{\Npone,\Szz\!-\!\frac{1}{2}}\rbrace \\
    \lbrace\ket{N,\Szz}\rbrace \\
    \lbrace\ket{\Npone,\Szz\!+\!\frac{1}{2}}\rbrace \\
    \lbrace\ket{N,\Szz\!+\!1}\rbrace
  \end{pmatrix},
\end{equation}
where each $\lbrace\ket{N,S_z}\rbrace$ corresponds to a vector of individual states $\ket{N,S_z,q_\uparrow,q_\downarrow}$, where $q_\sigma$ labels the configuration of the spin-$\sigma$ particles. In the presence of interactions, these states are not true eigenstates of the Hamiltonian, but they provide a qualitative approximation to the excitation gaps in most cases. 

The number of excited states to include is determined through a consideration of the chemical potentials for transitions to or from the ground states with spin $S_z\pm\frac{1}{2}$ as described above. With this arrangement for the configurations, the selection rules take the block-matrix form
\begin{equation}\label{eq:Selrules}
  \begin{pmatrix}
    0 & W^{+\downarrow}_{\Szz-1} & & & \cdots & 0 \\
    W^{-\downarrow}_{\Szz-\frac{3}{2}} & 0 & W^{-\uparrow}_{\Szz-\frac{1}{2}} & & & \vdots \\
     & W^{+\uparrow}_{\Szz-1} & 0 & W^{+\downarrow}_{\Szz}  & & \\
     & & W^{-\downarrow}_{\Szz-\frac{1}{2}} & 0 & W^{-\uparrow}_{\Szz+\frac{1}{2}} & \\
     \vdots & & & W^{+\uparrow}_{\Szz} & 0 & W^{+\downarrow}_{\Szz+1} \\
     0 & \cdots & & & W^{-\downarrow}_{\Szz+\frac{1}{2}} & 0 \\
  \end{pmatrix},
\end{equation}
where, assuming the vectors of states $\lbrace\ket{N,S_z}\rbrace$ are listed as subsequent groups of \spinup\ states (labeled by $q_\uparrow$) for each \spindn\ state (labeled by $q_\downarrow$), the sub-matrices $W^{\pm\sigma}_{S_z}$ are given by
\begin{subequations}
\begin{align}
  W^{\pm\uparrow}_{S_z} & = \mathbf{1}_\downarrow \otimes M^{\pm\uparrow}_{S_z}, \\
  W^{\pm\downarrow}_{S_z} & = M^{\pm\downarrow}_{S_z} \otimes
  \mathbf{1}_\uparrow.
\end{align}
\end{subequations}
The matrices $M^{\pm\sigma}_{S_z}$ contain the selection rules for transitions in the spin-$\sigma$ configuration individually, and are easily worked out by comparing the occupation vectors $\mathbf{n}_\sigma$ of the initial and final states. For example, $M^{+\uparrow}_{ij}=1$ whenever the $q_\uparrow=i$ state of the $\Npone$ configurations results from adding a single \spinup\ particle to the $q_\uparrow=j$ state of the $N$ configurations, which we can write as
\begin{equation}
  M^{+\uparrow}_{ij} = \begin{cases}
    1\quad\text{if}\quad
    \mathbf{n}_\uparrow(i)\cdot[\mathbf{1}-\mathbf{n}_\uparrow(j)]=1,\\
    0\quad\text{otherwise}.
  \end{cases}
\end{equation}
Similar relations determine the selection rules for other types of processes.

The rate matrix has a similar form to \eqnref{eq:Selrules}, where the nonzero selection rules are replaced by the transition rates
\begin{equation}
  R^{\pm\sigma}_{ij} =
  \sum_{r=\mathrm{S,D}}\Gamma^r_\sigma(\mu_{ij})W^{\pm\sigma}_{ij}f_r^\pm(\mu_{ij}),
\end{equation}
where $f^+_r = f_r$ is the Fermi function of lead $r$, and $f^-_r = 1-f_r$. As described in Sec. 2.4 in the main text, we then add diagonal elements to the rate matrix to impose a net balance of rates in equilibrium, and an extra row of ones to enforce normalization, constructing the master equation in the form of Eq. (10) in Ref. \cite{poteshman2019network}. The solution to this master equation gives the steady state occupation probability of each state $\ket{N,S_z,q_\uparrow,q_\downarrow}$, which we then use to compute the current flowing through the system. The current is most easily computed using \eqnref{eq:Ir1}, by isolating the transition rate involving only a single lead, e.g.\ for the source,
\begin{equation}
  S^{\pm\sigma}_{ij}=\GammaS_{\sigma}(\mu_{ij})W^{\pm\sigma}_{ij}f_\mathrm{S}^\pm(\mu_{ij}).
\end{equation}
Including signs to account for the direction of current flow, we can then write
\begin{equation}
  I = e\sum_{ij}T_{ij}P_j,
\end{equation}
where $P_j$ are the equilibrium occupation probabilities and
\begin{equation}
  T = \begin{pmatrix}
    0 & S^{+\downarrow}_{\Szz-1} & & & \cdots & 0 \\
    -S^{-\downarrow}_{\Szz-\frac{3}{2}} & 0 & -S^{-\uparrow}_{\Szz-\frac{1}{2}} & & & \vdots \\
     & S^{+\uparrow}_{\Szz-1} & 0 & S^{+\downarrow}_{\Szz}  & & \\
     & & -S^{-\downarrow}_{\Szz-\frac{1}{2}} & 0 & -S^{-\uparrow}_{\Szz+\frac{1}{2}} & \\
     \vdots & & & S^{+\uparrow}_{\Szz} & 0 & S^{+\downarrow}_{\Szz+1} \\
     0 & \cdots & & & -S^{-\downarrow}_{\Szz+\frac{1}{2}} & 0 \\
  \end{pmatrix}.\label{eq:Tmatrix}
\end{equation}
The block-diagonal form of Eq.~(\ref{eq:Tmatrix}) also facilitates the straightforward calculation of spin-resolved current components, simply by isolating the terms that correspond to tunneling of each spin species. The spin-resolved conductance is calculated as in Eq.~(\ref{eq:Conductance}) from a finite difference of the currents computed at two different settings for $\Vd$. 

The procedure can be generalized to account for additional effects. For example, we can include spin-conserving relaxation of excited states within each set $\lbrace\ket{N,S_z}\rbrace$ by adding block matrices describing these processes to the main diagonal of \eqnref{eq:Selrules}. Spin non-conserving relaxation due to spin-orbit coupling or the hyperfine interaction could also be included by adding terms to the next off-diagonal blocks (connecting states $\lbrace\ket{N,S_z}\rbrace$ with
$\lbrace\ket{N,S_z\!\pm\!1}\rbrace$). Note, however, that this model only obtains the steady-state ($t\rightarrow\infty$) configuration, and it assumes a Markovian (\textit{i.e.}, history-independent) interaction with the reservoirs; thus, we are not able to investigate coherent effects due to quantum evolution with this procedure.

We iterate this procedure until convergence is reached, adding additional $S_z$-configurations and excited states until the occupation probability of the outermost states falls below a predetermined threshold. In the results reported here and in the main text, the total occupation probability threshold of the outermost states was constrained to be less than 0.02. When performing simulations over a range of different parameters, the matrix-construction procedure is followed independently for each bias setting, and therefore the set of states included in the calculation varies. When investigating the role of some network measures, it is important to maintain a constant network size (\textit{i.e.}, the total number of states) and ideally the same state definitions. Therefore, after performing the calculation once over the full parameter space of interest, we determine the union of all quantum states that appear at any point and subsequently repeat the calculation at each bias point using the full set. A drawback of this approach is that, at every bias point, a large number of states have negligible occupation probability and do not contribute to the dynamics. When appropriate, the non-participatory states can be removed using a thresholding procedure as described in the next section.

\subsection{Calculating current and conductance\label{sec:CalcIG}}

Once the master equation has been solved for the probabilities $P(s)$, we can compute the current flowing out of each lead from the expression
\begin{equation} \label{eq:Ir1}
\begin{split}
  I_r & = e\sum_{s\sp} \bigl[ \gamma^+_{r,\sptos}P(\sp)
      - \gamma^-_{r,s\rightarrow\sp}P(s)\bigr] \\
      & = e\sum_{s\sp}\bigl[\gamma^+_{r,\sptos} -
      \gamma^-_{r,\sptos}\bigr] P(\sp),
\end{split}
\end{equation}
where we have used the single-particle Hamiltonian, given by Equation 1 in Ref \cite{poteshman2019network} to simplify the second term. Using the relation
\begin{equation}
  \sum_r\bigl[\gamma^+_{r,\sptos} -
      \gamma^-_{r,\sptos}\bigr] = \Bigl(N(s)-N(\sp)\Bigr)\gamma_{\sp
      s},
\end{equation}
it is straightforward to show that $\sum_r I_r = 0$, i.e.\ that the total current is conserved. Dropping the dependence on $\ket{\ls}$, we can write \eqnref{eq:Ir1} in the form
\begin{equation}\label{eq:Ir2}
  I_r = e\sum_{s\sp}\Gamma_r(\mussp) M_{s\sp}\Bigl[f_r(\mussp)P(\sp)
  - \bigl(1-f_r(\mussp)\bigr)P(s)\Bigr],
\end{equation}
where
\begin{equation}
  M_{s\sp} = \sum_\ls
  \bigl\lvert\bra{s}a^\dag_\ls\ket{\sp}\bigr\rvert^2
\end{equation}
represents the selection rules for transitions between states
$s^\prime\leftrightarrow s$ and $\mussp = E_s - E_\sp$ is the chemical potential for the antidot transition. Using current conservation we can derive the relation
\begin{equation}
  \sum_{s\sp}M_{s\sp}P(s) = \sum_{s\sp
  r}\frac{\Gamma_r}{\Gamma}M_{s\sp}\bigl[ P(\sp)+P(s)\bigr]
  f_r(\mussp),
\end{equation}
where $\Gamma = \sum_r \Gamma_r$ and we have suppressed the dependence of the $\Gamma$'s on $\mussp$. We use this relation to eliminate the term independent of $f_r$ in \eqnref{eq:Ir2} to obtain a final expression for the current out of lead $r$,
\begin{equation}
  I_r = e\sum_{s\sp r^\prime}
  \frac{\Gamma_r\Gamma_{r^\prime}}{\Gamma}
      M_{s\sp}\bigl[P(\sp)+P(s)\bigr]
      \times\bigl[f_r(\mussp) - f_{r^\prime}(\mussp)\bigr].
\end{equation}
We use this expression to calculate the current transmitted through the antidot, and compute the conductance at finite bias by
\begin{equation}\label{eq:Conductance}
  G(\Vd) = \frac{I(\Vd + \delta\Vd) - I(\Vd -
  \delta\Vd)}{2\delta\Vd},
\end{equation}
which is typically a good approximation if $e\delta\Vd\lesssim kT$.

\subsection{Thresholding the probability vector}

The rate matrix inversion calculations in our model of sequential transport through an antidot produced numerical inaccuracies in probability values near zero. To exclude these numerical inaccuracies, we found a threshold value that excludes small positive and negative values resulting from inaccuracies while maintaining the connectedness of the networks. Furthermore, we sought a low enough threshold that the sum of non-thresholded probability values remains close to 1.

The threshold of $1 \times 10^{-8}$ we used for all networks preserved both of these constraints; the probability vectors summed to greater than $0.999999$ for all networks corresponding to voltage settings where the magnitude of current was greater than $1 \si{\pico\ampere}$. This threshold maintained connectedness for all networks constructed using the effective model both including and excluding spin-conserving relaxation effects. For the networks constructed using the single-particle model, the threshold maintained connectedness for $0.9984 \%$ of networks excluding spin-conserving relaxation effects and for $0.9975 \%$ including spin-conserving relaxation effects.



\section{Supplementary Results}
\label{sec:suppl_results}

In this section we report the results of additional computations and analyses that complement those that were reported in the main manuscript.

\begin{figure*}
\centering
  \includegraphics[width=\textwidth]{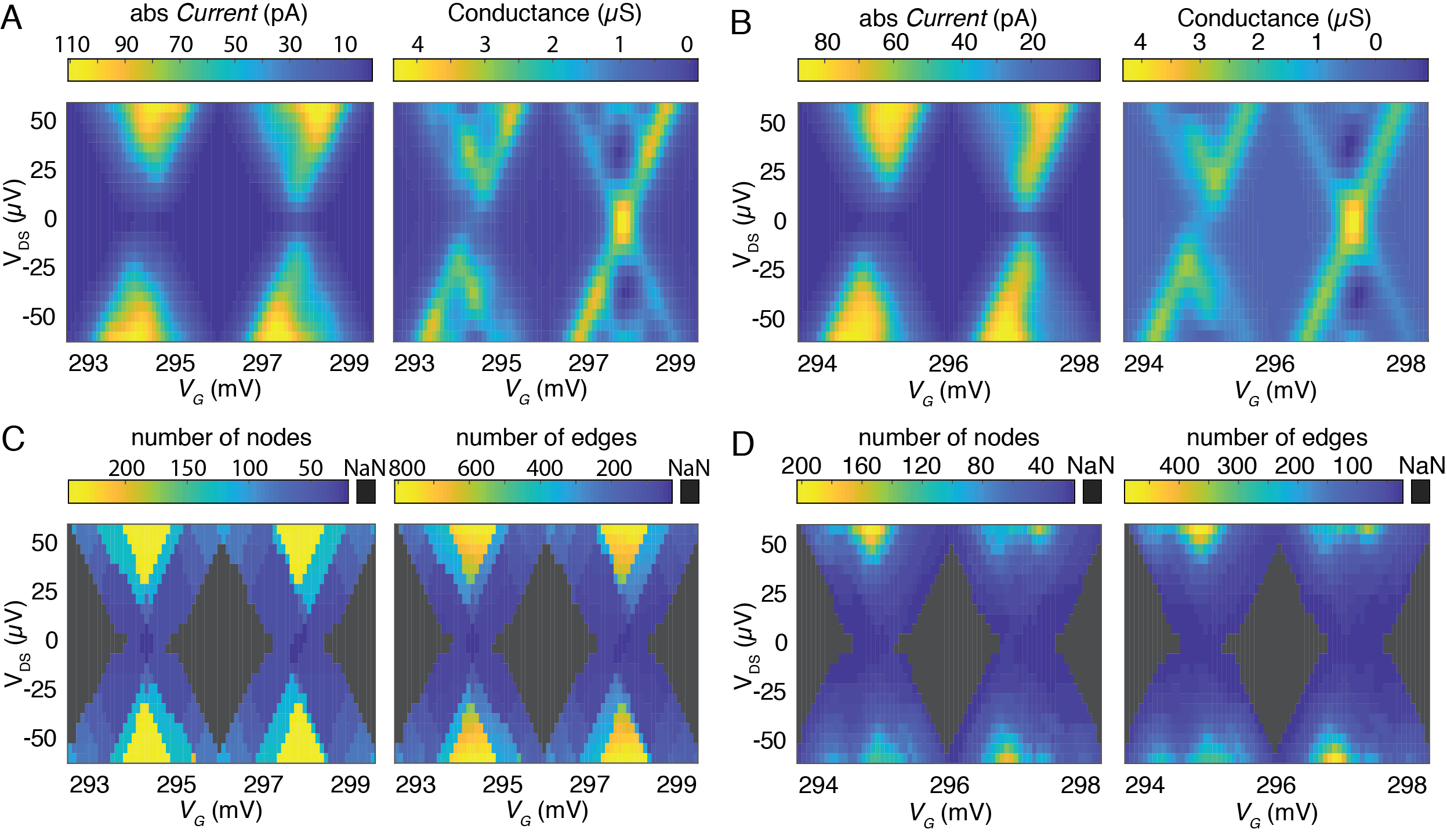}
  \caption{\textbf{Single-particle model versus effective model of non-equilibrium transport through quantum antidots with spin-conserving relaxations.} Current and conductance calculations based on a single-particle model \textbf{A.} and an effective model \textbf{B.} of energy states in quantum antidots including spin-conserving relaxation effects. The number of nodes and the number of edges for the networks corresponding to each set of voltage settings constructed from a single-particle model \textbf{C.} and an effective model \textbf{D.} of energy states including spin-conserving relaxation effects. Both models were run with the following parameters: T = $50$ \si{\milli\kelvin}, \textbf{B} = $1.2$ \si{\tesla}, effective spin-up tunneling rate $\Gamma_{\uparrow} = 0.005$ \si{\hertz}, effective spin-down tunneling rate $\Gamma_{\downarrow} = 0.0005$ \si{\hertz}, and relaxation rate = $\SI{0.005}{\hertz}$. All figures for simulations including spin-conserving relaxation effects are drawn from models run with these parameters unless noted otherwise. Note that we excluded networks corresponding to voltage settings that result in a current with magnitude less than $1$ \si{\pico\ampere} from our analysis, and values in panels \textbf{C.} and \textbf{D.} displayed as NaN indicate that networks were excluded.}
  \label{fig:CurrCond_relax}
\end{figure*}

\subsection{Current, conductance, and network size for the single-particle model and the effective model with spin-conserving relaxation effects}
\label{subsection:Curr_cond_relax}

We found that including spin-conserving relaxation effects in both models of non-equilibrium transport preserved the qualitative features of current and conductance (see Figure \ref{fig:CurrCond_relax}A-B), even though spin-conserving relaxation effects increase the number of edges in a network by allowing transitions between nodes with the same total number of particles (that is, transitions between two nodes with $N$ particles is allowed through a relaxation event). In the single-particle model, the spin-conserving relaxations increased the number of edges in a network by a factor of around 2 and increased the current through the antidot for a given set of voltage settings, whereas the increase in the number of edges for the effective model is a much smaller effect (see Figure \ref{fig:CurrCond_relax}C \& D). From a rudimentary analysis of network size (defined by the number of nodes and edges), we observe that the different mechanisms by which internal spin-conserving relaxations are encoded in the two models result in varying network architecture and thus may have different impacts on network functions such as transport and information storage.

\begin{figure*}
\centering
  \includegraphics[width=\textwidth]{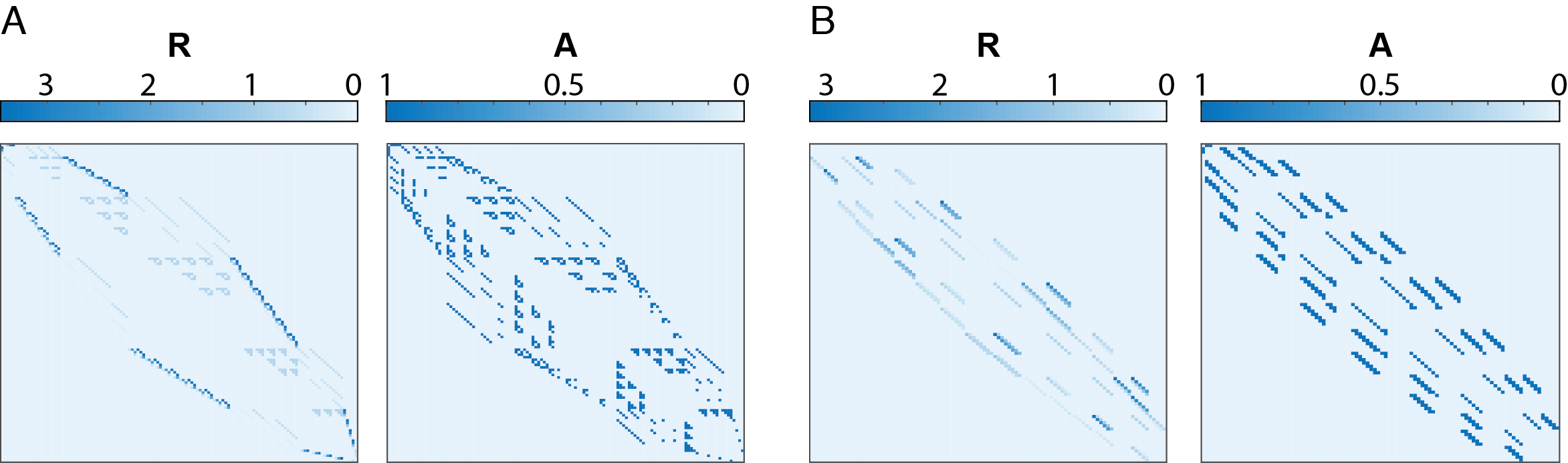}
  \caption{\textbf{Example rate and adjacency matrices including spin-conserving relaxation effects.} \textbf{A.} Example transition rate matrix \textbf{R} (left) and its corresponding adjacency matrix \textbf{A} (right) obtained using a single-particle model of quantum transport including spin-conserving relaxation effects. The network these matrices represent corresponds to voltage settings resulting in a current of $\lvert I \rvert$ = $81.0$ \si{\pico\ampere} and 112 states. \textbf{B.} Example transition rate matrix \textbf{R} (left) and its corresponding adjacency matrix \textbf{A} (right) obtained using an effective model of quantum transport including spin-conserving relaxation effects. The network that these matrices represent corresponds to voltage settings resulting in a current of $\lvert I \rvert$ = $81.8$ \si{\pico\ampere} and 98 states. }
  \label{fig:AdjMat_relax}
\end{figure*}

\subsection{Adjacency \& transition rate matrices with spin-conserving relaxation effects}
\label{subsection:adj_relax}

We included spin-conserving relaxation effects as described in Section 2.5 of the main text. We observe the relaxations in the rate matrix as the non-zero elements in the main-diagonal block matrices (see the left panels of Figure \ref{fig:AdjMat_relax}A-B). The rate matrices are then symmetrized and binarized, yielding the adjacency matrices on which we performed the network analyses (see the right panels of Figure \ref{fig:AdjMat_relax}). From the rate matrices, we can see the asymmetric nature of relaxations from an excited state to a ground state.

\begin{figure*}
\centering
  \includegraphics[width=\textwidth]{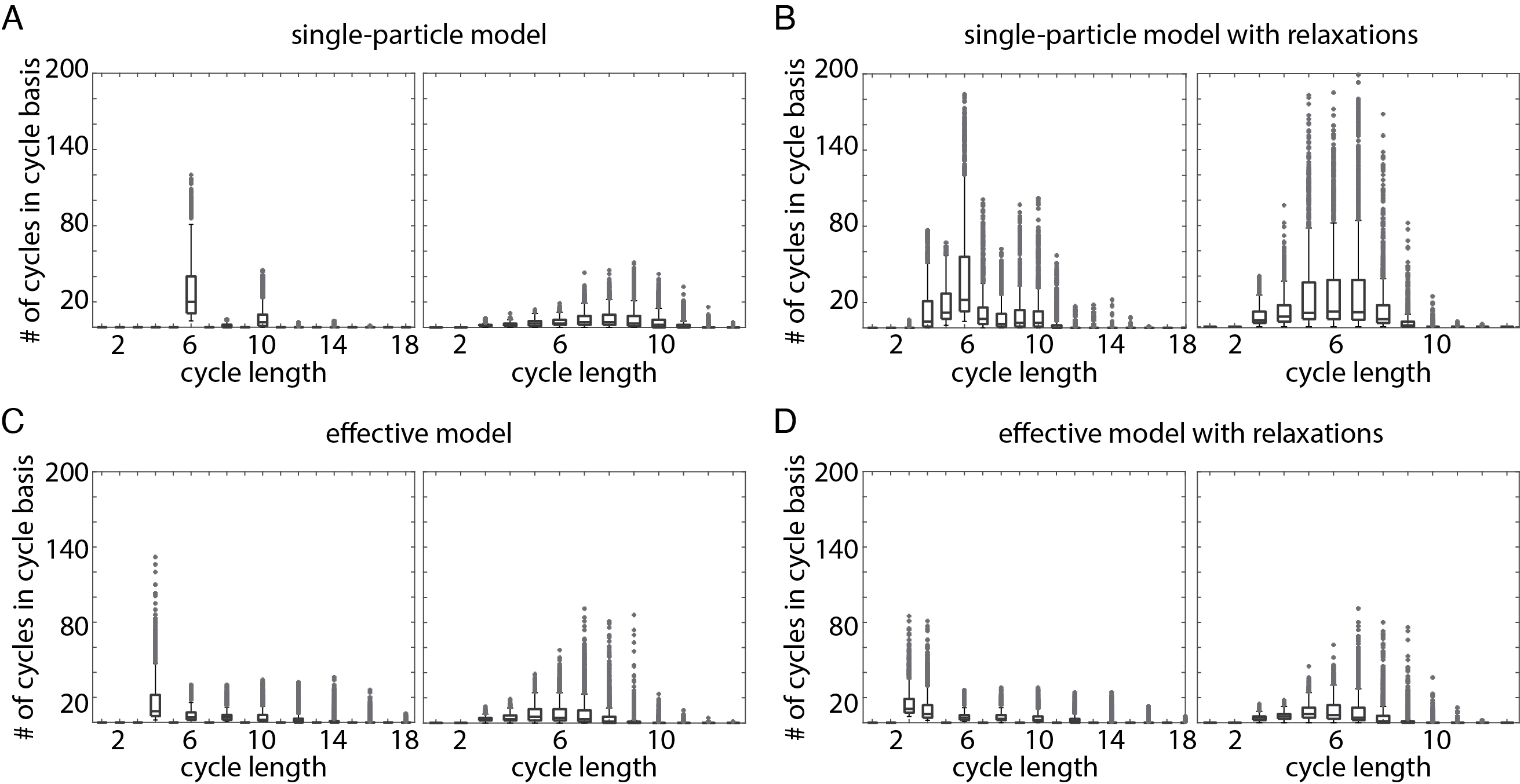}
  \caption{\textbf{Degree distribution and cycle structure are different aspects of network structure.} Distribution of cycle lengths in a cycle basis for all networks corresponding to a current greater than $1$ \si{\pico\ampere} for the networks (left) and the corresponding degree-distribution preserving random models (right) constructed from the single-particle model excluding spin-conserving relaxation effects \textbf{A.}, the single-particle model including spin-conserving relaxation effects \textbf{B.}, the effective model excluding spin-conserving relaxation effects \textbf{C.}, and the effective model including spin-conserving relaxation effects \textbf{D.}.}
  \label{fig:rand_models}
\end{figure*}

\subsection{Degree distribution \& cycle structure different phenomenon through random models}
\label{subsection:rand_models}

Both degree distribution and cycle structure are network properties that emerge from underlying quantum mechanical selection rules that govern mesoscopic quantum systems. To assess whether the cycle structure is dependent upon the degree distribution, we constructed degree-distribution preserving random networks based on the quantum transport networks \cite{maslov2002specificity, rubinov2010complex}. We found that the cycle structure is not dependent on the degree distribution, as the length distribution of cycle basis elements is different in the degree-distribution preserving random models (see Figure \ref{fig:rand_models}) compared with the quantum transport networks. For instance, all of the cycle bases from the random models include cycles of even length, so the bipartite nature of the networks constructed from models excluding spin-conserving relaxations (see the left panels of Figure \ref{fig:rand_models}A \& C) is a constraint that cannot be observed directly from the degree distribution. The independence of cycle structure and degree distribution in these quantum transport networks indicates that both may be important to take into consideration when designing network-based control strategies for quantum systems, given that both degree distribution and cycle structure play a role in the control profiles of complex networks \cite{campbell2015topological}.

\section{References}
\bibliography{ref_supp}
\bibliographystyle{unsrt}

%% file: LCB_macros.tex
\providecommand{\abs}[1]{\lvert#1\rvert}%

\providecommand{\ket}[1]{\lvert#1\rangle}

\def\HSP{\hat{h}}

\def\G2t{G_\mathrm{2T}}

\def\Szz{S_{z0}}

%% file: main.bbl
\begin{thebibliography}{10}

\bibitem{biamonte2019complex}
Jacob Biamonte, Mauro Faccin, and Manlio De~Domenico.
\newblock Complex networks from classical to quantum.
\newblock {\em Communications Physics}, 2(1):1--10, 2019.

\bibitem{lohe2010quantum}
Max~A. Lohe.
\newblock Quantum synchronization over quantum networks.
\newblock {\em Journal of Physics A: Mathematical and Theoretical},
  43(46):465301, 2010.

\bibitem{mulken2016complex}
Oliver M{\"u}lken, Maxim Dolgushev, and Mircea Galiceanu.
\newblock Complex quantum networks: From universal breakdown to optimal
  transport.
\newblock {\em Physical Review E}, 93(2):022304, 2016.

\bibitem{cabot2018unveiling}
Albert Cabot, Fernando Galve, V{\'\i}ctor~M Egu{\'\i}luz, Konstantin Klemm,
  Sabrina Maniscalco, and Roberta Zambrini.
\newblock Unveiling noiseless clusters in complex quantum networks.
\newblock {\em npj Quantum Information}, 4(1):1--9, 2018.

\bibitem{chakraborty2016spatial}
Shantanav Chakraborty, Leonardo Novo, Andris Ambainis, and Yasser Omar.
\newblock Spatial search by quantum walk is optimal for almost all graphs.
\newblock {\em Physical review letters}, 116(10):100501, 2016.

\bibitem{cirac1997quantum}
Juan~Ignacio Cirac, Peter Zoller, H~Jeff Kimble, and Hideo Mabuchi.
\newblock Quantum state transfer and entanglement distribution among distant
  nodes in a quantum network.
\newblock {\em Physical Review Letters}, 78(16):3221, 1997.

\bibitem{poteshman2019network}
Abigail~N Poteshman, Evelyn Tang, Lia Papadopoulos, Danielle~S Bassett, and
  Lee~C Bassett.
\newblock Network architecture of energy landscapes in mesoscopic quantum
  systems.
\newblock {\em New Journal of Physics}, 21(12):123049, 2019.

\bibitem{loss1998quantum}
Daniel Loss and David~P DiVincenzo.
\newblock Quantum computation with quantum dots.
\newblock {\em Physical Review A}, 57(1):120, 1998.

\bibitem{awschalom2013quantumspintronics}
David~D. Awschalom, Lee~C. Bassett, Andrew~S. Dzurak, Evelyn~L. Hu, and
  Jason~R. Petta.
\newblock Quantum spintronics: Engineering and manipulating atom-like spins in
  semiconductors.
\newblock {\em Science}, 339(6124):1174--1179, 2013.

\bibitem{barthelemy2013quantum}
Pierre Barthelemy and Lieven~MK Vandersypen.
\newblock Quantum dot systems: a versatile platform for quantum simulations.
\newblock {\em Annalen der Physik}, 525(10-11):808--826, 2013.

\bibitem{kouwenhoven1997introduction}
Leo~P Kouwenhoven, Gerd Sch{\"o}n, and Lydia~L Sohn.
\newblock Introduction to mesoscopic electron transport.
\newblock In {\em Mesoscopic Electron Transport}, pages 1--44. Springer, 1997.

\bibitem{wiel2002electron}
W.~G. van~der Wiel, S.~De~Franceschi, J.~M. Elzerman, T.~Fujisawa, S.~Tarucha,
  and L.~P. Kouwenhoven.
\newblock Electron transport through double quantum dots.
\newblock {\em Rev. Mod. Phys.}, 75:1--22, Dec 2002.

\bibitem{hanson2007spins}
Ronald Hanson, Leo~P Kouwenhoven, Jason~R Petta, Seigo Tarucha, and Lieven~MK
  Vandersypen.
\newblock Spins in few-electron quantum dots.
\newblock {\em Reviews of modern physics}, 79(4):1217, 2007.

\bibitem{sim2008electron}
H-S Sim, Masaya Kataoka, and Chris~JB Ford.
\newblock Electron interactions in an antidot in the integer quantum hall
  regime.
\newblock {\em Physics reports}, 456(4):127--165, 2008.

\bibitem{tighineanu2015unraveling}
Petru Tighineanu, Anders~S{\o}ndberg S{\o}rensen, S{\o}ren Stobbe, and Peter
  Lodahl.
\newblock Unraveling the mesoscopic character of quantum dots in nanophotonics.
\newblock {\em Physical review letters}, 114(24):247401, 2015.

\bibitem{yang2016resonance}
Chun-Jie Yang and Jun-Hong An.
\newblock Resonance fluorescence beyond the dipole approximation of a quantum
  dot in a plasmonic nanostructure.
\newblock {\em Physical Review A}, 93(5):053803, 2016.

\bibitem{bagrov2020detecting}
Andrey~A Bagrov, Mikhail Danilov, Sergey Brener, Malte Harland, Alexander~I
  Lichtenstein, and Mikhail~I Katsnelson.
\newblock Detecting quantum critical points in the $t- t'$ {Fermi-Hubbard}
  model via complex network theory.
\newblock {\em Scientific reports}, 10(1):1--9, 2020.

\bibitem{valdez2017quantifying}
Marc~Andrew Valdez, Daniel Jaschke, David~L Vargas, and Lincoln~D Carr.
\newblock Quantifying complexity in quantum phase transitions via mutual
  information complex networks.
\newblock {\em Physical Review Letters}, 119(22):225301, 2017.

\bibitem{sundar2018complex}
Bhuvanesh Sundar, Marc~Andrew Valdez, Lincoln~D Carr, and Kaden~RA Hazzard.
\newblock Complex-network description of thermal quantum states in the ising
  spin chain.
\newblock {\em Physical Review A}, 97(5):052320, 2018.

\bibitem{zaman2019real}
Shehtab Zaman and Wei-Cheng Lee.
\newblock Real-space visualization of quantum phase transitions by network
  topology.
\newblock {\em Physical Review E}, 100(1):012304, 2019.

\bibitem{andrieux2006fluctuation}
David Andrieux and Pierre Gaspard.
\newblock Fluctuation theorem for transport in mesoscopic systems.
\newblock {\em Journal of Statistical Mechanics: Theory and Experiment},
  2006(01):P01011, 2006.

\bibitem{andrieux2007fluctuation}
David Andrieux and Pierre Gaspard.
\newblock Fluctuation theorem for currents and schnakenberg network theory.
\newblock {\em Journal of statistical physics}, 127(1):107--131, 2007.

\bibitem{stone1991edge}
Michael Stone.
\newblock Edge waves in the quantum hall effect.
\newblock {\em Annals of Physics}, 207(1):38--52, 1991.

\bibitem{stone1990schur}
Michael Stone.
\newblock Schur functions, chiral bosons, and the quantum-hall-effect edge
  states.
\newblock {\em Physical Review B}, 42(13):8399, 1990.

\bibitem{bassett2019probing}
Lee~C Bassett.
\newblock Probing electron-electron interactions with a quantum antidot.
\newblock {\em arXiv:1912.08006}, 2019.

\bibitem{lizier2012information}
Joseph~T Lizier, Fatihcan~M Atay, and J{\"u}rgen Jost.
\newblock Information storage, loop motifs, and clustered structure in complex
  networks.
\newblock {\em Physical Review E}, 86(2):026110, 2012.

\bibitem{boccaletti2006complex}
Stefano Boccaletti, Vito Latora, Yamir Moreno, Martin Chavez, and D-U Hwang.
\newblock Complex networks: Structure and dynamics.
\newblock {\em Physics reports}, 424(4-5):175--308, 2006.

\bibitem{campbell2015topological}
Colin Campbell, Justin Ruths, Derek Ruths, Katriona Shea, and R{\'e}ka Albert.
\newblock Topological constraints on network control profiles.
\newblock {\em Scientific reports}, 5:18693, 2015.

\bibitem{mace1995general}
DR~Mace, CHW Barnes, G~Faini, D~Mailly, MY~Simmons, CJB Ford, and M~Pepper.
\newblock General picture of quantum hall transitions in quantum antidots.
\newblock {\em Physical Review B}, 52(12):R8672, 1995.

\bibitem{stone1992edge}
Michael Stone, HW~Wyld, and Roy~L. Schult.
\newblock Edge waves in the quantum hall effect and quantum dots.
\newblock {\em Physical Review B}, 45(24):14156, 1992.

\bibitem{macdonald1993quantum}
Allan~H. MacDonald, SR~Eric Yang, and Michael~D. Johnson.
\newblock Quantum dots in strong magnetic fields: Stability criteria for the
  maximum density droplet.
\newblock {\em Australian Journal of Physics}, 46(3):345--358, 1993.

\bibitem{ashcroft1976solid}
Neil~W Ashcroft and N~David Mermin.
\newblock Solid state physics, 1976.

\bibitem{Luttinger1963}
Joaquin~M. Luttinger.
\newblock An exactly soluble model of a many-fermion system.
\newblock {\em J. Math. Phys.}, 4(9):1154--1162, 1963.

\bibitem{Tomonaga1950}
Shinichiro Tomonaga.
\newblock Remarks on bloch's method of sound waves applied to many-fermion
  problems.
\newblock {\em Prog. Theor. Phys.}, 5:544, 1950.

\bibitem{newman2010networks}
Mark Newman.
\newblock {\em Networks: An Introduction}.
\newblock Oxford University Press, 2010.

\bibitem{albert2002statistical}
R{\'e}ka Albert and Albert-L{\'a}szl{\'o} Barab{\'a}si.
\newblock Statistical mechanics of complex networks.
\newblock {\em Reviews of modern physics}, 74(1):47, 2002.

\bibitem{lynn2020human}
Christopher~W Lynn, Lia Papadopoulos, Ari~E Kahn, and Danielle~S Bassett.
\newblock Human information processing in complex networks.
\newblock {\em Nature Physics}, 16(9):965--973, 2020.

\bibitem{kim2018role}
Jason~Z Kim, Jonathan~M Soffer, Ari~E Kahn, Jean~M Vettel, Fabio Pasqualetti,
  and Danielle~S Bassett.
\newblock Role of graph architecture in controlling dynamical networks with
  applications to neural systems.
\newblock {\em Nature physics}, 14(1):91--98, 2018.

\bibitem{lydon2020hunters}
David~M Lydon-Staley, Dale Zhou, Ann~Sizemore Blevins, Perry Zurn, and
  Danielle~S Bassett.
\newblock Hunters, busybodies and the knowledge network building associated
  with deprivation curiosity.
\newblock {\em Nature human behaviour}, pages 1--10, 2020.

\bibitem{sizemore2018knowledge}
Ann~E Sizemore, Elisabeth~A Karuza, Chad Giusti, and Danielle~S Bassett.
\newblock Knowledge gaps in the early growth of semantic feature networks.
\newblock {\em Nature human behaviour}, 2(9):682--692, 2018.

\bibitem{sizemore2018cliques}
Ann~E Sizemore, Chad Giusti, Ari Kahn, Jean~M Vettel, Richard~F Betzel, and
  Danielle~S Bassett.
\newblock Cliques and cavities in the human connectome.
\newblock {\em Journal of computational neuroscience}, 44(1):115--145, 2018.

\bibitem{garcia2012building}
Guadalupe~Clara Garcia, Annick Lesne, Marc-Thorsten H{\"u}tt, and Claus~C
  Hilgetag.
\newblock Building blocks of self-sustained activity in a simple deterministic
  model of excitable neural networks.
\newblock {\em Frontiers in computational neuroscience}, 6:50, 2012.

\bibitem{garcia2014role}
Guadalupe~Clara Garcia, Annick Lesne, Claus~C Hilgetag, and Marc-Thorsten
  H{\"u}tt.
\newblock Role of long cycles in excitable dynamics on graphs.
\newblock {\em Physical Review E}, 90(5):052805, 2014.

\bibitem{bianconi2003number}
Ginestra Bianconi and Andrea Capocci.
\newblock Number of loops of size h in growing scale-free networks.
\newblock {\em Physical review letters}, 90(7):078701, 2003.

\bibitem{hagberg2008exploring}
Aric Hagberg, Pieter Swart, and Daniel S~Chult.
\newblock Exploring network structure, dynamics, and function using networkx.
\newblock Technical report, Los Alamos National Lab.(LANL), Los Alamos, NM
  (United States), 2008.

\bibitem{paton1969algorithm}
Keith Paton.
\newblock An algorithm for finding a fundamental set of cycles of a graph.
\newblock {\em Communications of the ACM}, 12(9):514--518, 1969.

\bibitem{katifori2010damage}
Eleni Katifori, Gergely~J Sz{\"o}ll{\H{o}}si, and Marcelo~O Magnasco.
\newblock Damage and fluctuations induce loops in optimal transport networks.
\newblock {\em Physical review letters}, 104(4):048704, 2010.

\bibitem{kavitha2009cycle}
Telikepalli Kavitha, Christian Liebchen, Kurt Mehlhorn, Dimitrios Michail,
  Romeo Rizzi, Torsten Ueckerdt, and Katharina~A Zweig.
\newblock Cycle bases in graphs characterization, algorithms, complexity, and
  applications.
\newblock {\em Computer Science Review}, 3(4):199--243, 2009.

\bibitem{safar2009counting}
Maytham Safar, Khalid Alenzi, and Saud Albehairy.
\newblock Counting cycles in an undirected graph using dfs-xor algorithm.
\newblock In {\em 2009 First International Conference on Networked Digital
  Technologies}, pages 132--139. IEEE, 2009.

\bibitem{guillaume2004bipartite}
Jean-Loup Guillaume and Matthieu Latapy.
\newblock Bipartite structure of all complex networks.
\newblock {\em Information processing letters}, 90(5):215--221, 2004.

\bibitem{newman2001random}
Mark~EJ Newman, Steven~H Strogatz, and Duncan~J Watts.
\newblock Random graphs with arbitrary degree distributions and their
  applications.
\newblock {\em Physical review E}, 64(2):026118, 2001.

\bibitem{barabasi2003scale}
Albert-L{\'a}szl{\'o} Barab{\'a}si and Eric Bonabeau.
\newblock Scale-free networks.
\newblock {\em Scientific American}, 288(5):60--69, 2003.

\bibitem{amaral2000classes}
Luis~A Amaral, Antonio Scala, Marc Barthelemy, and H~Eugene Stanley.
\newblock Classes of small-world networks.
\newblock {\em Proceedings of the national academy of sciences},
  97(21):11149--11152, 2000.

\bibitem{baggio2020data}
Giacomo Baggio, Danielle~S Bassett, and Fabio Pasqualetti.
\newblock Data-driven control of complex networks.
\newblock {\em arXiv preprint arXiv:2003.12189}, 2020.

\bibitem{zanudo2017structure}
Jorge Gomez~Tejeda Za{\~n}udo, Gang Yang, and R{\'e}ka Albert.
\newblock Structure-based control of complex networks with nonlinear dynamics.
\newblock {\em Proceedings of the National Academy of Sciences},
  114(28):7234--7239, 2017.

\bibitem{motter2015networkcontrology}
Adilson~E Motter.
\newblock Networkcontrology.
\newblock {\em Chaos: An Interdisciplinary Journal of Nonlinear Science},
  25(9):097621, 2015.

\bibitem{nacher2013structural}
Jose~C Nacher and Tatsuya Akutsu.
\newblock Structural controllability of unidirectional bipartite networks.
\newblock {\em Scientific reports}, 3:1647, 2013.

\bibitem{daley2014quantum}
Andrew~J Daley.
\newblock Quantum trajectories and open many-body quantum systems.
\newblock {\em Advances in Physics}, 63(2):77--149, 2014.

\bibitem{ramakrishna1996relation}
Viswanath Ramakrishna and Herschel Rabitz.
\newblock Relation between quantum computing and quantum controllability.
\newblock {\em Physical Review A}, 54(2):1715, 1996.

\bibitem{lind2005cycles}
Pedro~G Lind, Marta~C Gonzalez, and Hans~J Herrmann.
\newblock Cycles and clustering in bipartite networks.
\newblock {\em Physical review E}, 72(5):056127, 2005.

\bibitem{zhou2018cycle}
Xiaoping Zhou, Xun Liang, Jichao Zhao, and Shusen Zhang.
\newblock Cycle based network centrality.
\newblock {\em Scientific reports}, 8(1):1--11, 2018.

\bibitem{bohn2007structure}
Steffen Bohn and Marcelo~O Magnasco.
\newblock Structure, scaling, and phase transition in the optimal transport
  network.
\newblock {\em Physical review letters}, 98(8):088702, 2007.

\bibitem{ma2008ordered}
Avi Ma'ayan, Guillermo~A Cecchi, John Wagner, A~Ravi Rao, Ravi Iyengar, and
  Gustavo Stolovitzky.
\newblock Ordered cyclic motifs contribute to dynamic stability in biological
  and engineered networks.
\newblock {\em Proceedings of the National Academy of Sciences},
  105(49):19235--19240, 2008.

\bibitem{woodhouse2016stochastic}
Francis~G Woodhouse, Aden Forrow, Joanna~B Fawcett, and J{\"o}rn Dunkel.
\newblock Stochastic cycle selection in active flow networks.
\newblock {\em Proceedings of the National Academy of Sciences},
  113(29):8200--8205, 2016.

\bibitem{liu2011controllability}
Yang-Yu Liu, Jean-Jacques Slotine, and Albert-L{\'a}szl{\'o} Barab{\'a}si.
\newblock Controllability of complex networks.
\newblock {\em nature}, 473(7346):167--173, 2011.

\bibitem{tanizawa2005optimization}
Toshi Tanizawa, Gerald Paul, Reuven Cohen, Shlomo Havlin, and H~Eugene Stanley.
\newblock Optimization of network robustness to waves of targeted and random
  attacks.
\newblock {\em Physical review E}, 71(4):047101, 2005.

\bibitem{tarjan1972depth}
Robert Tarjan.
\newblock Depth-first search and linear graph algorithms.
\newblock {\em SIAM journal on computing}, 1(2):146--160, 1972.

\bibitem{chung2019exact}
Moo~K Chung, Hyekyoung Lee, Alex DiChristofano, Hernando Ombao, and Victor
  Solo.
\newblock Exact topological inference of the resting-state brain networks in
  twins.
\newblock {\em Network Neuroscience}, 3(3):674--694, 2019.

\bibitem{sizemore2019importance}
Ann~E Sizemore, Jennifer~E Phillips-Cremins, Robert Ghrist, and Danielle~S
  Bassett.
\newblock The importance of the whole: topological data analysis for the
  network neuroscientist.
\newblock {\em Network Neuroscience}, 3(3):656--673, 2019.

\bibitem{blevins2020topology}
Ann~Sizemore Blevins and Danielle~S Bassett.
\newblock Topology in biology.
\newblock {\em Journal: Handbook of the Mathematics of the Arts and Sciences},
  pages 1--23, 2020.

\bibitem{bianconi2005loops}
Ginestra Bianconi and Matteo Marsili.
\newblock Loops of any size and hamilton cycles in random scale-free networks.
\newblock {\em Journal of Statistical Mechanics: Theory and Experiment},
  2005(06):P06005, 2005.

\end{thebibliography}


\begin{thebibliography}{1}

\bibitem{bassett2019probing}
Lee~C Bassett.
\newblock Probing electron-electron interactions with a quantum antidot.
\newblock {\em arXiv preprint arXiv:1912.08006}, 2019.

\bibitem{bethe1986intermediate}
H.~A. Bethe and R.~W. Jackiw.
\newblock {\em Intermediate Quantum Mechanics}.
\newblock Benjamin Cummings, Menlo Park, California, 1986.

\bibitem{jackiw2018intermediate}
Roman Jackiw.
\newblock {\em Intermediate quantum mechanics}.
\newblock CRC Press, 2018.

\bibitem{poteshman2019network}
Abigail~N Poteshman, Evelyn Tang, Lia Papadopoulos, Danielle~S Bassett, and
  Lee~C Bassett.
\newblock Network architecture of energy landscapes in mesoscopic quantum
  systems.
\newblock {\em New Journal of Physics}, 21(12):123049, 2019.

\bibitem{maslov2002specificity}
Sergei Maslov and Kim Sneppen.
\newblock Specificity and stability in topology of protein networks.
\newblock {\em Science}, 296(5569):910--913, 2002.

\bibitem{rubinov2010complex}
Mikail Rubinov and Olaf Sporns.
\newblock Complex network measures of brain connectivity: uses and
  interpretations.
\newblock {\em Neuroimage}, 52(3):1059--1069, 2010.

\bibitem{campbell2015topological}
Colin Campbell, Justin Ruths, Derek Ruths, Katriona Shea, and R{\'e}ka Albert.
\newblock Topological constraints on network control profiles.
\newblock {\em Scientific reports}, 5:18693, 2015.

\end{thebibliography}
